\documentclass[twocolumn]{aa} 

\usepackage{graphicx}
\usepackage{txfonts}
\begin{document} 
\title{Gravitational atmospheric tides as a probe of Titan's interior: Application to Dragonfly}

   \author{B. Charnay
          \inst{1}
          \and
          G. Tobie\inst{2}
          \and
          S. Lebonnois\inst{3}
          \and
          R. D. Lorenz\inst{4}   
          }

   \institute{LESIA, Observatoire de Paris, Universit\'e PSL, CNRS, Sorbonne Universit\'e, Universit\'e de Paris, 5 Place Jules Janssen, 92195 Meudon, France. \\
              \email{benjamin.charnay@obspm.fr}
         \and
             Laboratoire de Plan\'etologie de Nantes, Nantes, France
         \and
             Laboratoire de M\'et\'eorologie Dynamique, Paris, France
         \and
             Johns Hopkins  Applied Physics Lab, Laurel, MD, USA
             }


 
  \abstract
   {Saturn’s massive gravity is expected to causes a tide in Titan's atmosphere, producing a surface pressure variation through the orbit of Titan and tidal winds in the troposphere. The future Dragonfly mission could analyse this exotic meteorological phenomenon.} 
   {We analyse the effect of Saturn's tides on Titan's atmosphere and interior to determine how pressure measurements by Dragonfly could constrain Titan's interior.}
   {We model atmospheric tides with analytical calculations and with a 3D Global Climate Model (the IPSL-Titan GCM), including the tidal response of the interior.}
   {We predict that the Love numbers of Titan's interior should verify 1 + $\Re(k_2-h_2$) $\sim$ 0.02-0.1 and $\Im(k_2-h_2$) $<$ 0.04.
   The deformation of Titan’s interior should therefore strongly weaken gravitational atmospheric tides, yielding a residual surface pressure amplitude of only $\sim$ 5 Pa, with a phase shift of 5-20 hours.
     Tidal winds are very weak, of the order of 3$\times$10$^{-4}$ m s$^{-1}$ in the lower troposphere.  Finally, constraints from Dragonfly data may permit the real and the imaginary parts of $k_2-h_2$ to be estimated with a precision of $\pm$0.01-0.03.} 
   {Measurements of pressure variations by Dragonfly over the whole mission could give valuable constraints on the thickness of Titan's ice shell, and via geophysical models, its heat flux and the density of Titan's internal ocean.}

\keywords{Planets and satellites: individual: Titan - Planets and satellites: atmospheres - Planets and satellites: interiors}
\maketitle
%

\section{Introduction}

Just as Saturn's massive gravity causes tides both in Titan's interior as well as its surface seas, it causes a tide in the atmosphere \citep{lorenz1992, tokano2002, strobel2006}.
\cite{tokano2002} analysed these gravitational atmospheric tides with a 3D Global Climate Model (GCM). They found that the tidal potential produced by Saturn would generate a surface pressure variation of $\sim$1.3 hPa through the orbit of Titan and tidal winds in the troposphere, with a mean wind speed of $\sim$0.3-0.4 m/s at 300 m. \cite{walterscheid2006} and \cite{strobel2006} studied these gravitational tidal waves with an analytical model. They found that tidal winds should increase with altitude until saturation in the upper atmosphere, where they would deposit energy. \cite{walterscheid2006} suggested that the vertical transport by gravitational tides could produce the haze layers in Titan's upper atmosphere.
However, these atmospheric studies did not take into account the deformation of Titan's interior. Titan likely possesses an internal water-rich ocean, as suggested from the elevated Love number $k_2\sim$0.62 measured by the spacecraft Cassini \citep{iess2012, durante2019}. Such a large induced gravitational potential, implies a large deformation of the interior which should strongly impact atmospheric tides. Measuring them could provide additional constraints on Titan's internal structure.

Dragonfly, a relocatable lander for Titan \citep{lorenz2018}, is presently under development as NASA 4th New Frontiers mission, with a view to launch in 2027 and arrival in 2034.  Its objectives include assessing Titan's habitability, and thus probing Titan's internal structure. In this context, this paper indicates a new means by which Dragonfly may address this topic through measurements of pressure variations. 
In Section 2, we express the tidal potential exerted by Saturn on Titan's atmosphere, including the deformation of the interior. We reanalyse gravitational atmospheric tides with analytical calculations and with a 3D GCM in Section 3. Then, we discuss the possibility to constraint Titan's interior with Dragonfly in Section 4.
We finish with a summary and conclusions in Section 5.

\section{Computation of Saturn's tidal potential for Titan's atmosphere}

\subsection{Expression of Saturn's tidal potential}

Titan orbits around Saturn with a synchronous rotation rate. Its large orbital eccentricity ($e=0.0292$) leads to time-dependent tides. Tides are due both to the time variation of the distance between Titan and Saturn (radial tide) and to the time variation of Saturn's angular position from the subsaturnian point (librational tide). The tidal potential produced by Saturn can be expressed  as \citep{sagan1982, tokano2002}:

\begin{equation}
\begin{split} 
&V=-\frac{G M_{\rm S}}{a}\left(\frac{R_{\rm T}}{a}\right)^2 \left(\frac{3}{2} \cos^2(\lambda)\cos^2(\phi)-1 \right) -3 e\frac{ G M_{\rm S}}{a}\left(\frac{R_{\rm T}}{a}\right)^2  \times \\
&  \left[\frac{1}{2}(3\cos^2(\lambda)\cos^2(\phi)-1) \cos(\Omega t) + \cos^2(\lambda)\sin(2\phi)\sin(\Omega t) \right]
\end{split} 
\label{eq1}
\end{equation}
where $\phi$ is the longitude from the sub-saturnian point, $\lambda$ is the latitude, $e$ = 0.0292 is Titan's eccentricity around Saturn, $G$ = 6.673$\times$10$^{-11}$ N m$^2$ kg$^{-2}$ is the universal gravitational constant, $M_{\rm S}$ = 5.685$\times$10$^{26}$ kg is the mass of Saturn, $R_{\rm T}$ = 2575 km is Titan’s radius, $a$ = 1.222$\times$10$^9$ m is Titan's semi major axis, $t$ is the time (measured from periapsis, when Titan is at the nearest point of Saturn), $\Omega$=4.56$\times$10$^{-6}$ rad s$^{-1}$ is Titan's orbital angular velocity.

\begin{figure}[h!] 
\begin{center} 
\includegraphics[width=8cm,clip]{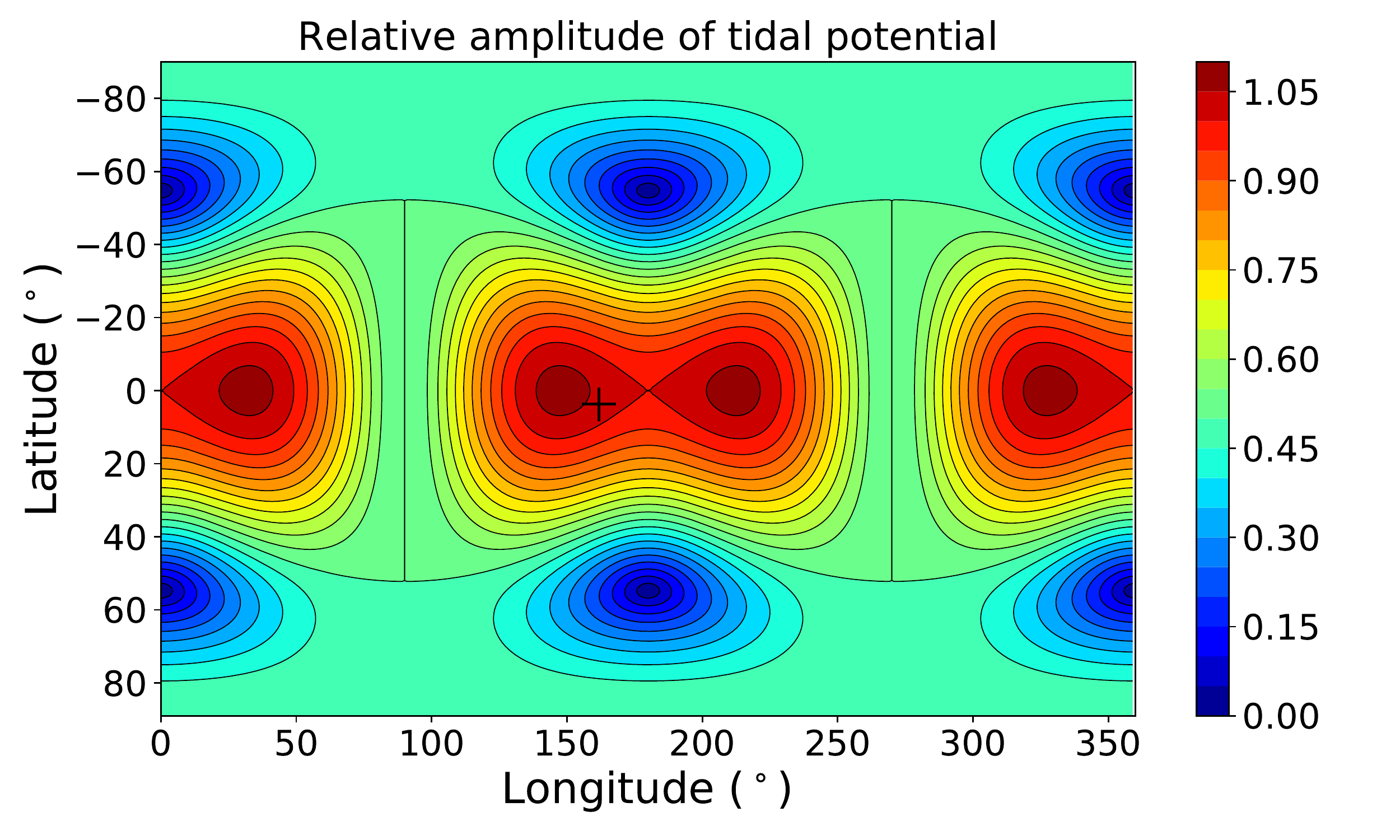} 
\includegraphics[width=8cm,clip]{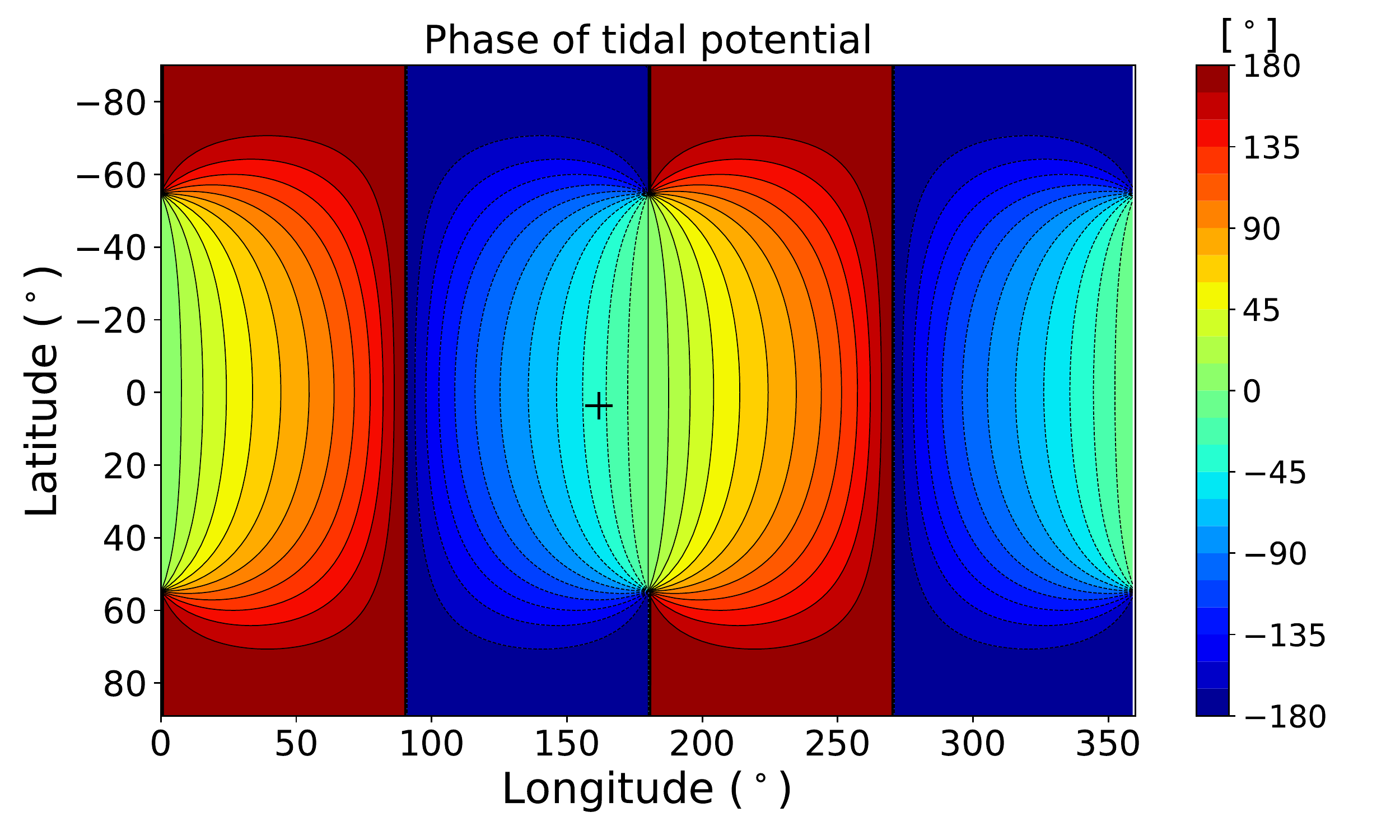} 
\end{center} 
\caption{Relative amplitude ($V_1/V_0$) and phase ($\psi$) of the dynamic tidal potential as a function of longitude and latitude. The black cross indicates Dragonfly's landing site.}
\label{figure_1}
\end{figure}

The first term in \eqref{eq1} is time independent. This implies a permanent modification of Titan's geoid and interior with no impact on the atmospheric dynamics. We focus on the time dependent terms of \eqref{eq1}, which are related to Titan's eccentricity. The dynamic tidal potential can be expressed as:

\begin{equation}
\begin{split} 
& V_{\rm dyn}= \frac{-V_0}{2}(3\cos^2(\lambda)\cos^2(\phi)-1) \cos( \Omega t) \\
& + V_0 \cos^2(\lambda)\sin(2\phi)\sin(\Omega t) 
\end{split} 
\label{eq2}
\end{equation}
with
\begin{equation}
V_0=-\frac{G M_S}{a}\left(\frac{R_T}{a}\right)^2 3\ e
\label{eq3}
\end{equation}
where $V_0$=-12.08 m$^2$ s$^{-2}$.
We can express this potential as a superposition of prograde (eastward), retrograde (westward) and stationary waves:
\begin{equation}
\begin{split} 
V_{\rm dyn}=-V_0 \left[ \frac{7}{8}\cos^2(\lambda)\cos(2 \phi - \Omega t) - \frac{1}{8}\cos^2(\lambda)\cos(2 \phi + \Omega t) \right] \\
+ V_0 \left(\frac{3}{4}\cos^2(\lambda)- \frac{1}{2} \right)\cos(\Omega t)
\end{split} 
\label{eq4}
\end{equation}
or with normalized associated Legendre polynomials $P_{r,s}(\mu)$ and $\mu=\sin\lambda$:
\begin{equation}
\begin{split} 
& V_{\rm dyn}=- \frac{7}{6}\sqrt{\frac{3}{5}} V_0 P_{2,2}(\mu) \cos(2 \phi - \Omega t) \\
& + \frac{1}{6}\sqrt{\frac{3}{5}} V_0 P_{2,2}(\mu) \cos(2 \phi + \Omega t)  - \frac{V_0}{\sqrt{10}} P_{2,0}(\mu)\cos(\Omega t)
\end{split} 
\label{eq5}
\end{equation}
The first term is a diurnal eastward wave (wavenumber=2), the second term is a diurnal westward wave (wavenumber=2) and the last term is a stationary wave. The eastward wave dominates at low latitudes and the stationary wave dominates at high latitudes.
We can also express the dynamic tidal potential \eqref{eq2} as:
\begin{equation}
V_{\rm dyn}=V_1 \cos(\psi-\Omega t) 
\label{eq6}
\end{equation}
with $V_1=-V_0 \sqrt{ \frac{1}{4}\left( 3\cos^2(\lambda) \cos^2(\phi)-1 \right)^2 + \cos^4(\lambda)\sin^2(2\phi)} $  and $\psi=\arctan\left[\frac{2 \cos^2(\lambda)\sin(2\phi)}{3\cos^2(\lambda)\cos^2(\phi) -1}\right]$.
This expression directly gives the local wave amplitude as shown in Fig. \ref{figure_1}. The tidal wave amplitude is maximal for $\lambda$ = 0$^\circ$ N and $\phi$ = $\pm$ 32$^\circ$ (modulo 180$^\circ$). It is null for $\lambda$ = $\pm$ 60$^\circ$ N and $\phi$ = 0$^\circ$ / 180$^\circ$. It is equal to $V_0$ at the subsaturnian and anti-saturnian points.

\cite{tyler2008} pointed out the presence of obliquity tidal forces on a moon with a non-zero obliquity. In such a case, the obliquity tidal potential is given by: 
\begin{equation}
V_{\rm obliquity}=\frac{3}{2}\Omega^2 R_T^2 \theta_0 \cos(\lambda) \sin(\lambda) \left(  \cos(\phi - \Omega t) + \cos(\phi + \Omega t) \right)
\end{equation}
where $\theta_0$ is the obliquity in radians. $\theta_0$=0.0052 rad for Titan (0.3$^\circ$). Using Kepler's third law ($\Omega^2=\frac{G M_S}{a^3}$), the obliquity tidal potential can be expressed as:
\begin{equation}
V_{\rm obliquity}=-V_0\frac{\theta_0}{2e}\cos(\lambda) \sin(\lambda) \left(  \cos(\phi - \Omega t) + \cos(\phi + \Omega t) \right)
\end{equation}
The ratio of the amplitude of the obliquity tide by the amplitude of the eccentricity tide is approximately $\frac{\theta_0}{2e}tan(\lambda)$. The obliquity tide dominates at latitudes higher than 80$^\circ$ on Titan. It is thus necessary to take into account the obliquity tide to study Titan’s sea tides (see \cite{tokano2010b,tokano2014,vincent2016,vincent2018}). In contrast, the eccentricity tide dominates at low latitudes. Its amplitude should be $\sim$100 times that of the obliquity tide at Dragonfly's landing site (7$^\circ$ N 199$^\circ$W). We therefore neglect the effect of the obliquity tide throughout this study, which focuses on Titan's equatorial regions.

\subsection{Influence of the deformation of Titan's interior}

The moment of inertia inferred from Cassini gravity measurements ($C/MR^2\simeq 0.34$, \cite{iess2010})  indicates that Titan's interior is differentiated into a outer hydrosphere and an inner rocky core \citep{castillo2010,sotin2021}. The relatively high value of $C/MR^2$ suggest a low density and large rocky core ($>2050-2100$ km), potentially containing a significant fraction of organics \citep{neri2020}. The determination of time variations of Titan's gravity field due to Saturn's tides by Cassini \citep{iess2012,durante2019}  provided a clear evidence that Titan possess a liquid layer underneath its icy surface, most likely a water ocean \citep{mitri2014}. The high value of the Love number, $k_2$ = 0.616 $\pm$ 0.067, which quantify the amplitude of induced-gravitational potential, suggested that the ocean is dense and salty \citep{mitri2014}. However, as the uncertainties on $k_2$ are still large, the ocean depth and density still remains unconstrained. Due to the high pressure reached at the bottom of the hydrosphere, high-pressure ice polymorphs (phase V and VI) are expected to form \citep[e.g.][]{journaux2020}. The thickness of this high-pressure layer is conditioned by the thickness of the outer ice shell and the composition of the ocean. For a thin outer layer ($<50-70$ km), it is possible that  no high-pressure ice layer exists and that the ocean is in direct contact with the rocky core, which would strongly enhance the astrobiological potential of Titan's internal ocean. Constraining the outer ice shell thickness would therefore provide crucial constraints on the habitability of Titan.

The deformation of Titan's interior in response to Saturn's tides results in an change of the gravitational potential  induced by time-varying internal mass redistribution, proportional to the gravitational Love number $k_2$ mentioned above, together with a time-varying deflection of the surface, proportional to the displacement Love number $h_2$. During high tides, $k_2$ results in an increase of tidal potential exerted on Titan's atmosphere and hence induced surface pressure, while $h_2$ results in a reduction due to surface elevation. If the interior behaves as perfectly elastic body, the internal response would be in phase with the tidal potential produced by Saturn (Eq.\ref{eq1}). In reality, Titan's interior is not perfectly elastic, part of the mechanical energy is dissipated in the interior  and surface by various friction processes \citep{sohl1995, tobie2005, tokano2014, lorenz2014}. This results in an out-of-phase term which amplitude depends on the mechanical properties of the interior. The in-phase term is represented by the real part of the Love number, whereas the out-of-phase term corresponds to the imaginary part.
As a consequence, the tidal potential exerted on Titan's atmosphere can be expressed as:
\begin{equation}
V_{\rm atm}=V_1 \left[(1+ \Re(k_2-h_2))\cos(\Omega t-\psi) - \Im(k_2-h_2)\sin(\Omega t-\psi) \right]
\label{eq7}
\end{equation}
where $k_2$ and $h_2$ are the 2nd-degree complex Love number. $\Re()$ and $\Im()$ are the real part of these complex numbers .
The local amplitude of the tidal potential is thus reduced by a factor $\gamma_2 = \sqrt{(1 + \Re(k_2-h_2))^2+\Im(k_2-h_2)^2}$ \citep{sohl1995, tokano2014, lorenz2014} and has a phase shift $\psi_2=-\arctan\left(\frac{\Im(k_2-h_2)}{1 + \Re(k_2-h_2)}\right)$. \\

Pre-Cassini interior models predicted that $\gamma_2$ should be small, lower than 0.2 and probably lower than 0.1 \citep[e.g.]{sohl2003}. Here we compute the complex Love numbers $k_2$ and $h_2$ following the approach used in \cite{mitri2014} assuming a viscoelastic compressible interior \citep{tobie2005} and considering interior structure models reproducing the mean density (1881 kg/m$3$) and the reduced moment of inertia (0.341) of Titan. We consider interior structures consisting of four main layers from center to surface: a rocky core, a high-pressure (HP) ice V–VI layer, a liquid water ocean, an ice I layer.
The tidal response is computed using the formulation of the spheroidal deformation developed by  \cite{takeuchi1972}, initially derived for the elastic case, extended to the viscoelastic case by solving it in the frequency domain and by defining complex shear and bulk modulii, equivalent to the elastic modulii used in the elastic problem (see \cite{tobie2005} for more details). For the liquid ocean layer, the simplified formulation of \cite{saito1974} is employed assuming a quasi-static and non-dissipative fluid material. For computing the viscoelastic deformation of the solid layers, we assume a compressible Andrade model, defined from bulk modulus, $K$, shear modulus, $S$, viscosity, $\eta$ and two empirical parameters $\alpha$ and $\beta$ describing the transient response of the viscoelastic medium \citep{castillo2011}. Following \cite{castillo2011}, we assume the relationship $\beta\sim S^{-(1-\alpha)}\eta^{-\alpha}$ and choose a reference value of 0.25 for $\alpha$. The rheological parameters are assumed constant in each internal layer, except in the outer ice layer where the viscosity, $\eta$, as a function of depth is computed from a given temperature profile, which is either conductive  or convective
 In the conductive case, a temperature profile varying linearly from the surface to the bottom of the ice shell is considered. In the convective case, the ice shell is separated in a conductive lid and a convective isothermal sublayer.  The thickness of the conductive lid, $b_{lid}$, is determined from an imposed surface heat flux and the viscosity of the convective layer is determined from Nusselt–Rayleigh scaling laws \citep{dumoulin1999} in order to reproduce a convective heat flux equal to the imposed surface heat flux, $\phi_S$, varying between 15 and 30 mW.m$^{-2}$ (see appendix A in \cite{lefevre2014} for more details). 
In the conductive part of the ice shell, the viscosity is computed from the linear temperature profile using an Arrhenius law:
\begin{equation}
    \eta(z)=\eta_{T_b}\exp\left(\frac{E_a}{R(T(z)-T_b)}\right)
\end{equation}
where $T_b$ is the temperature at the bottom of the conductive shell, $E_a$ is the activation energy ($E_a=50$ kJ.mol$^{-1}$), $R$ is the gas constant.

\begin{figure}[h!] 
\begin{center} 
\includegraphics[width=8cm,clip]{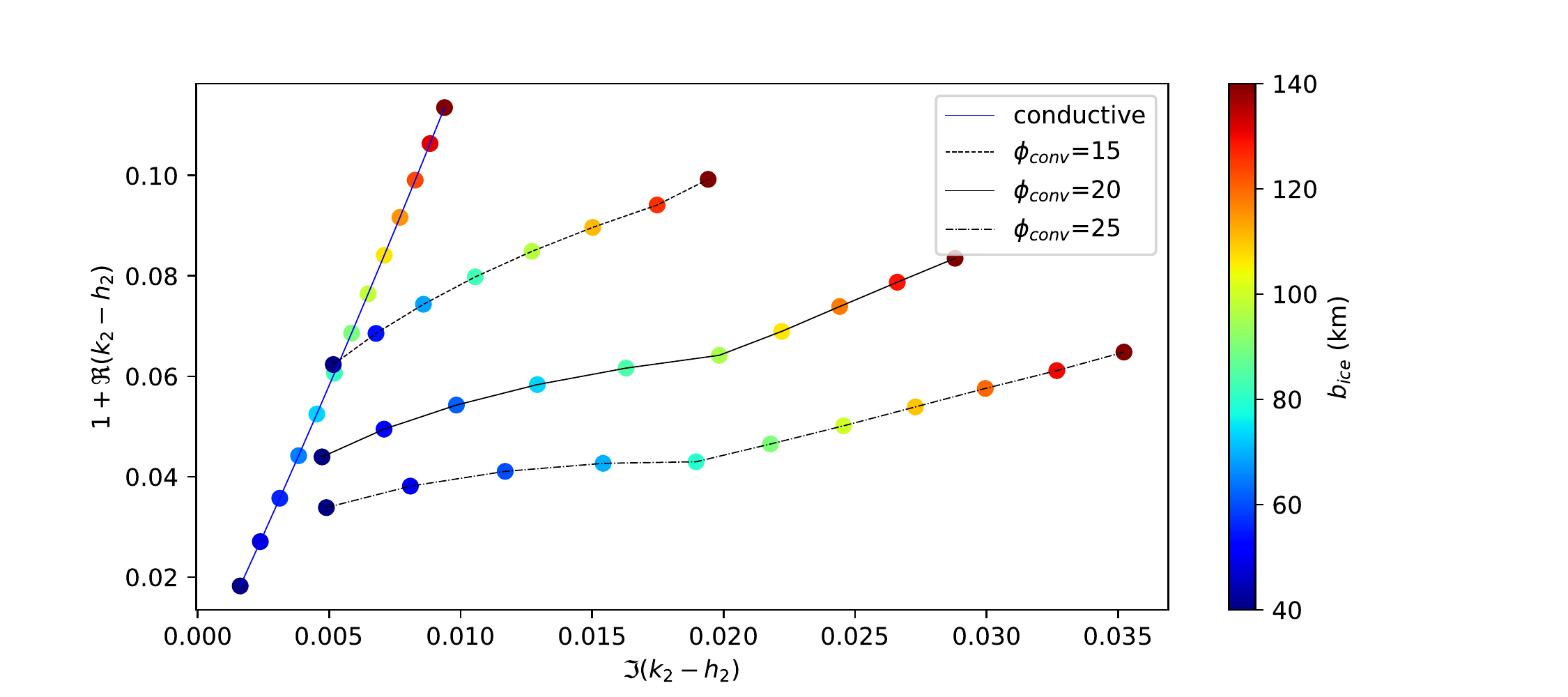} 
\includegraphics[width=8cm,clip]{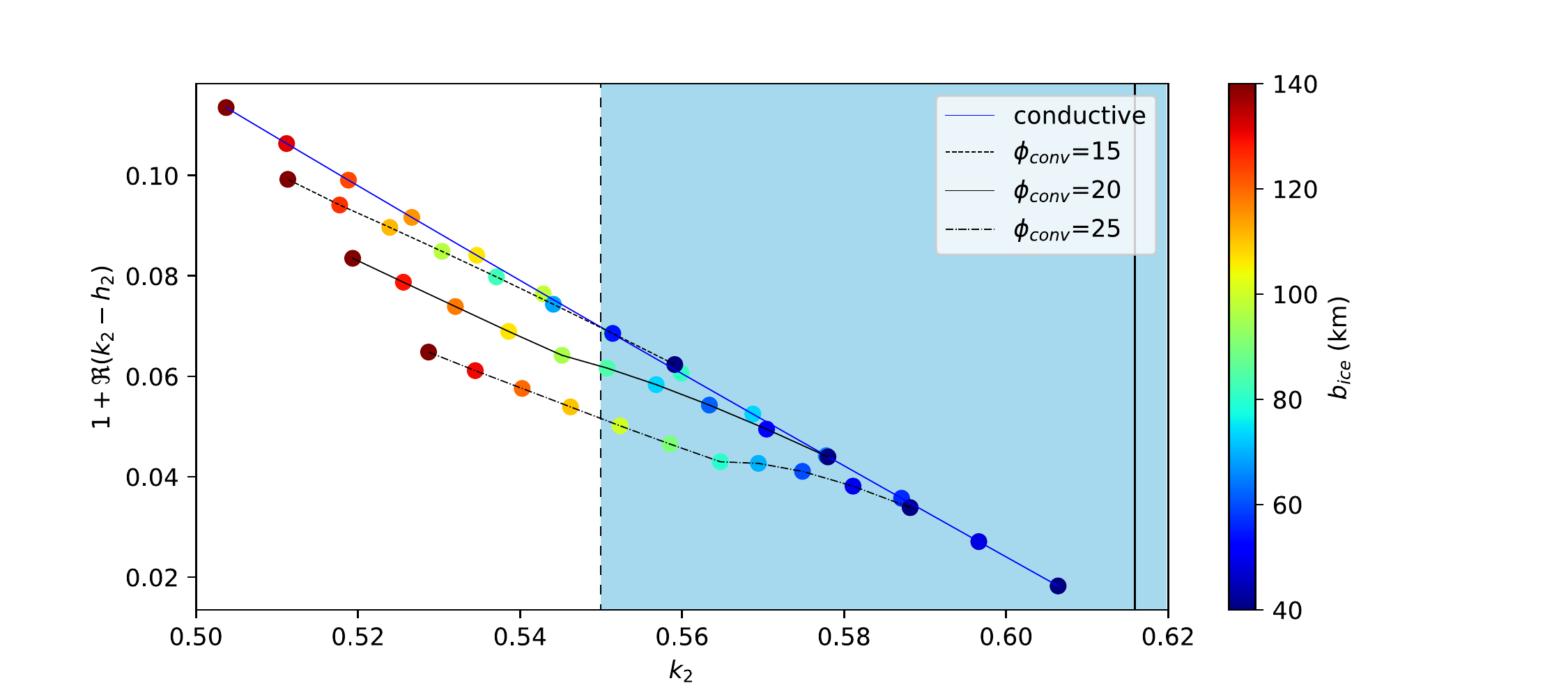} 
\includegraphics[width=8cm,clip]{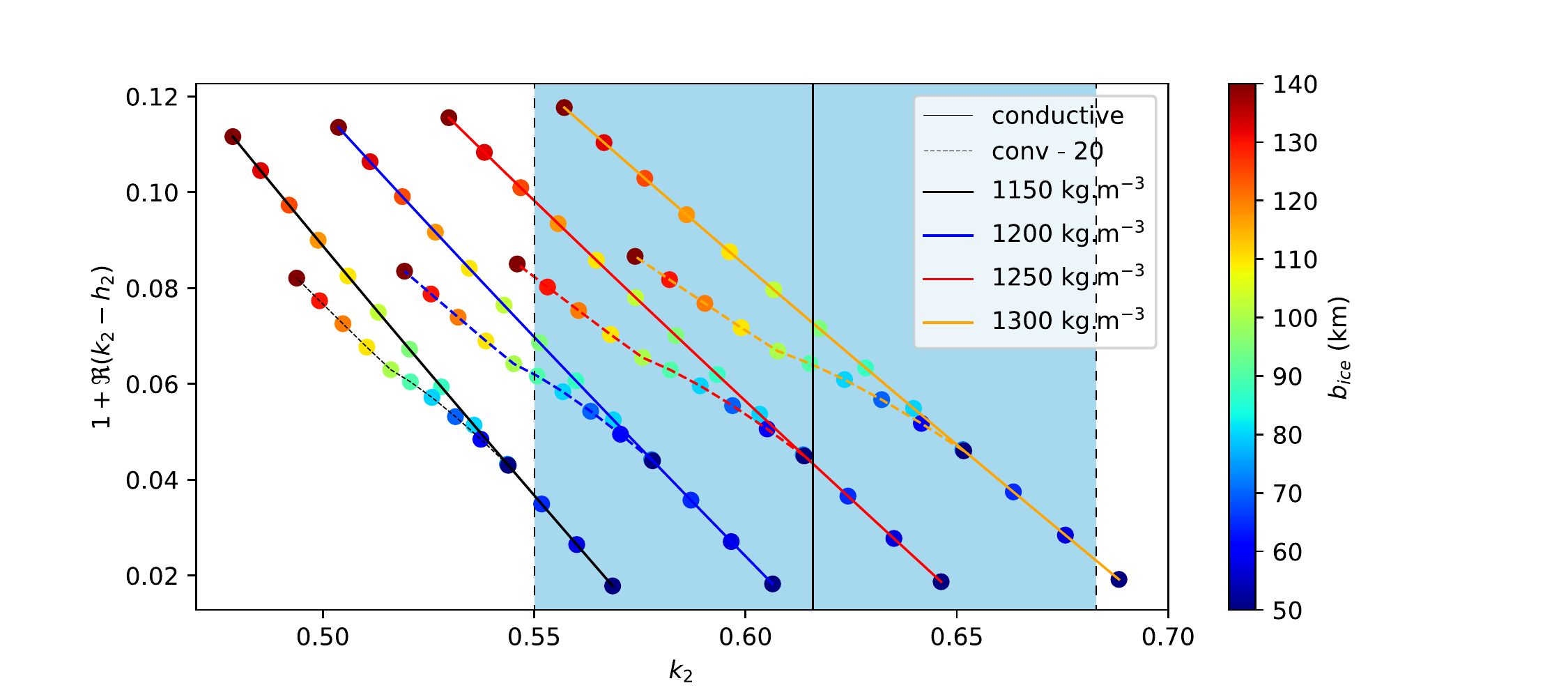}
\end{center} 
\caption{Titan's values of $1+\Re(k_2-h_2)$, $\Im(k_2-h_2)$ and $\Re(k_2)$, as a function of the ice shell thickness, the convective flux (given in mW/m$^2$) and the internal ocean density. The average ocean density was fixed at 1200 kg/m$^3$ for the top and the middle panel. The blue area in the middle and bottom panels shows the values compatible with Cassini data at 1-sigma ($\Re(k_2)$ = 0.549-0.683).}
\label{figure_2}
\end{figure}

Fig. \ref{figure_2} shows the computed values of $1+\Re(k_2-h_2)$, $\Im(k_2-h_2)$ and $\Re(k_2)$, as a function of the ice shell thickness, the convective heat flux and the ocean density. 
This figure suggests that accurate measurements of $1+\Re(k_2-h_2)$ and $\Im(k_2-h_2)$ could constrain the ice shell thickness, the thermal state of the ice shell (conductive vs. convective) and potentially give some estimate of the internal heat flux. For a conductive ice shell, 
$\Re(\gamma_2)=1+\Re(k_2-h_2)$ is directly proportional to the ice shell thickness. For a convective ice shell, the relationship between $\Re(\gamma_2)$ and the ice shell thickness is less straightforward. However, by assessing both $\Re(\gamma_2)$ and $\Im(\gamma_2)$, the ice shell thickness and the vigor of convection can be determined unambiguously.  
Once the ice shell thickness constrained from $\gamma_2$ measurements, the measurements of $\Re(k_2)$ can be used to constrain the density of the ocean and hence its solute content.

\begin{figure*}[h!] 
\begin{center} 
\includegraphics[width=8cm,clip]{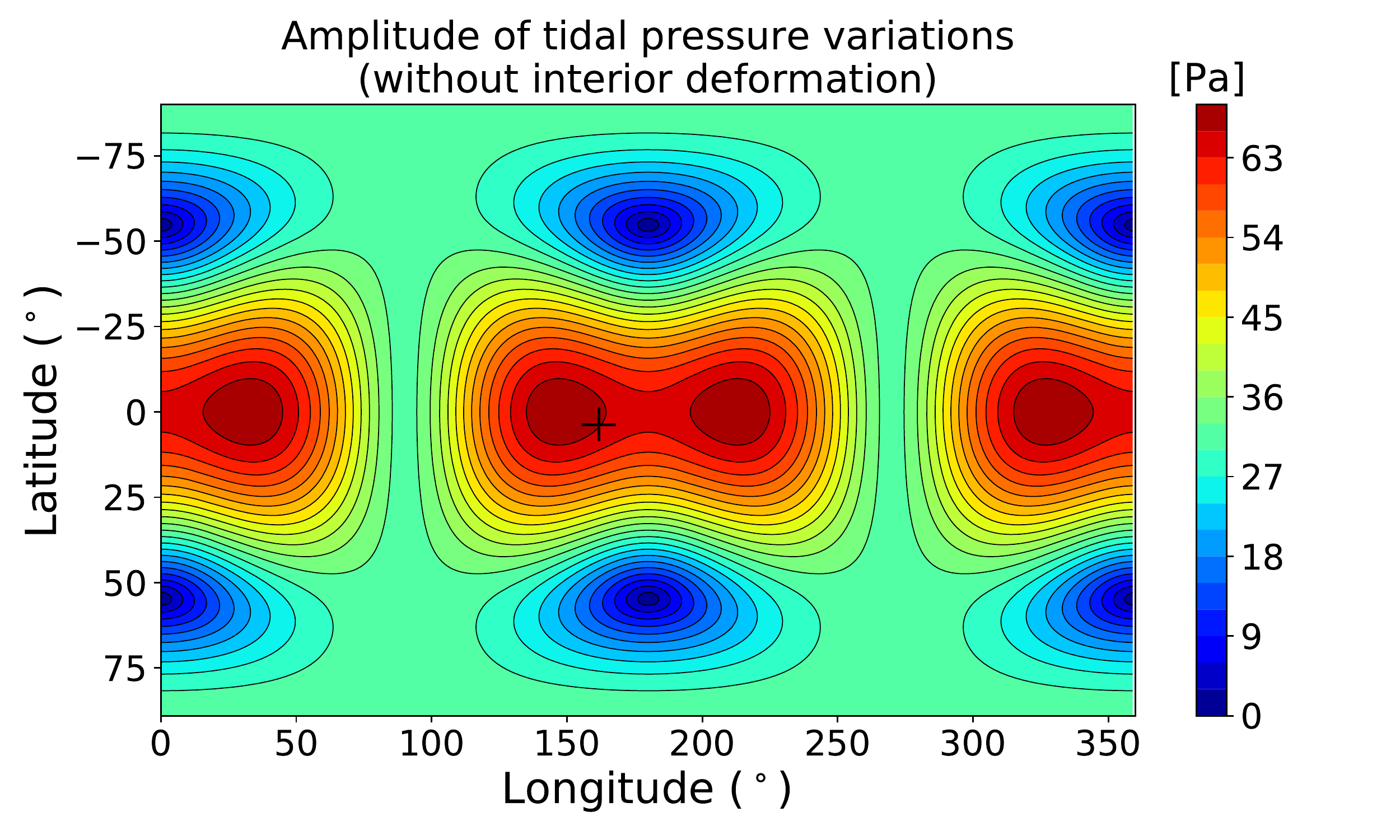} 
\includegraphics[width=8cm,clip]{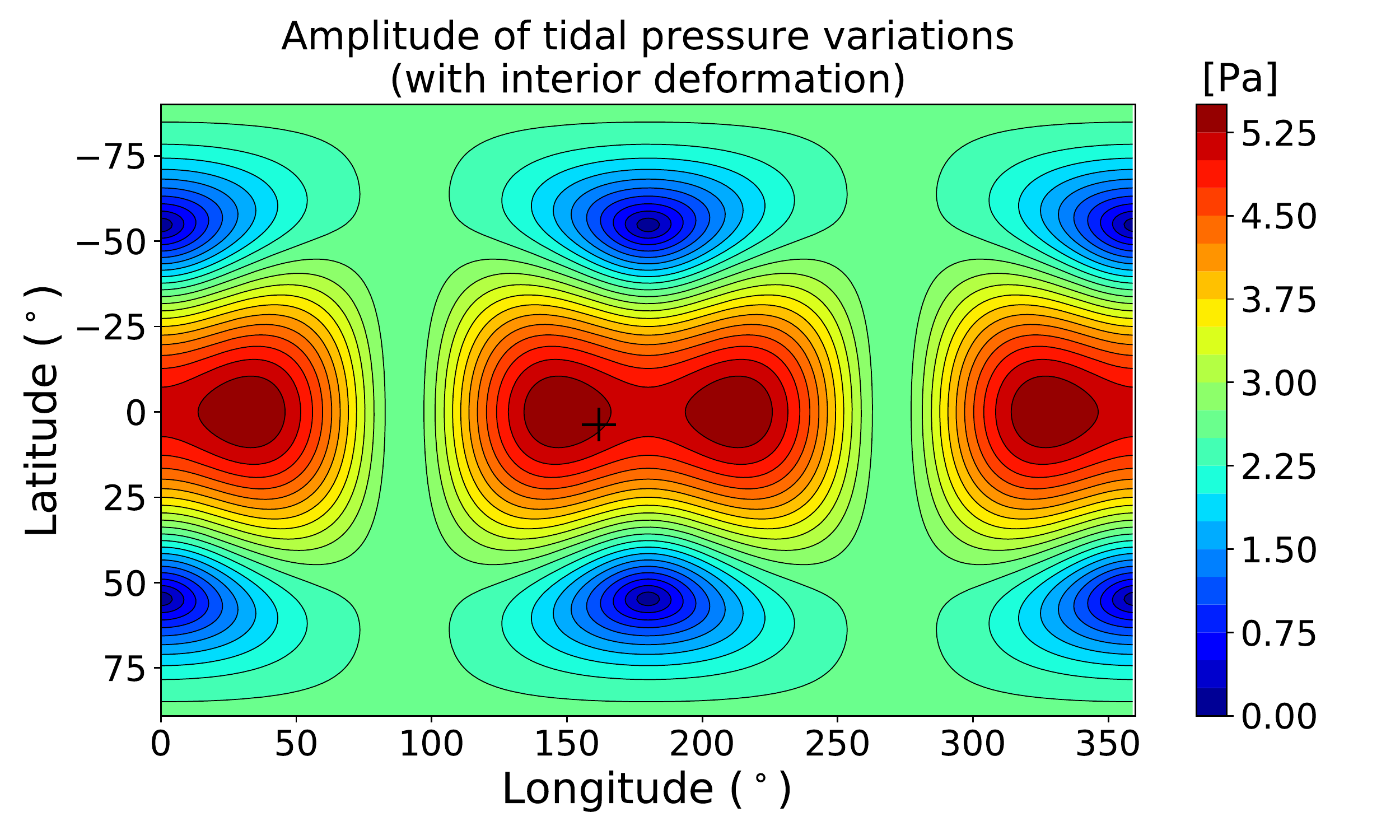} 
\includegraphics[width=8cm,clip]{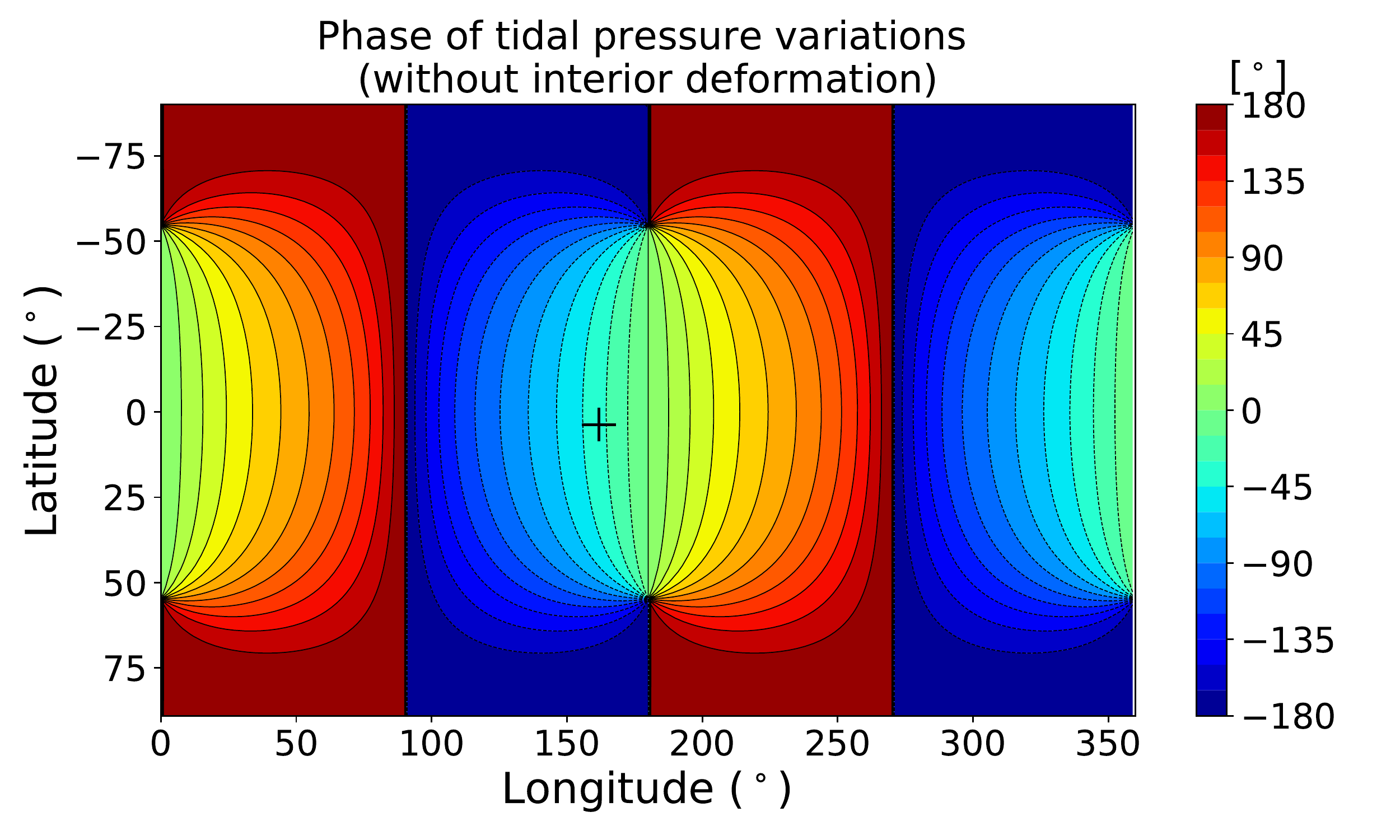}
\includegraphics[width=8cm,clip]{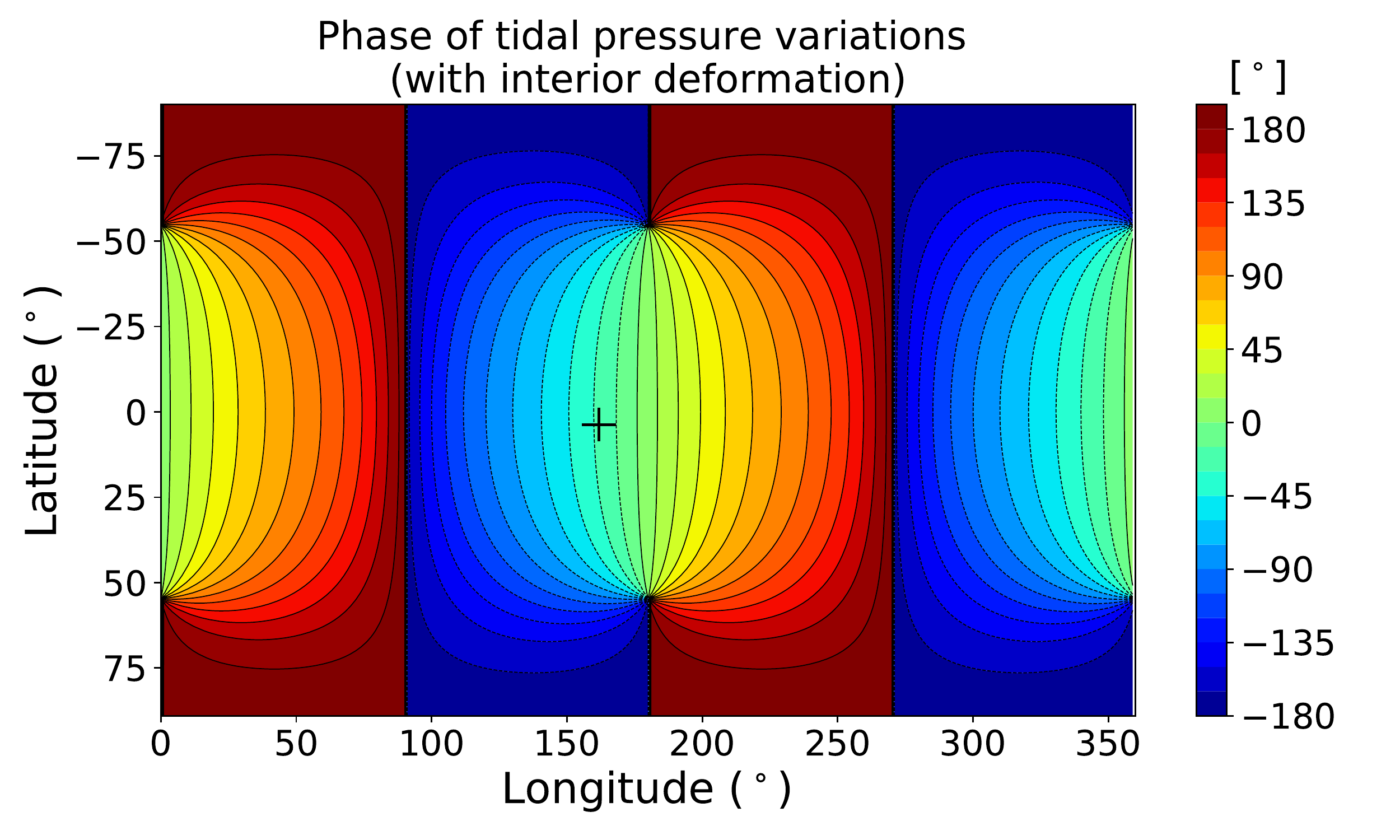} 
\end{center} 
\caption{Amplitude of the tidal pressure variations (\textit{top panel}) and phase (\textit{bottom panel}) as a function of longitude and latitude, without interior deformation (\textit{left}) and with interior deformation (\textit{right}). The latter is computed with 1+$\Re(k_2-h_2$)=0.08 and $\Im(k_2-h_2$)=0.01. The black cross indicates Dragonfly's landing site.}
\label{figure_3}
\end{figure*}

As demonstrated in the next section, the gradient of surface pressure perfectly compensate the gradient of tidal potential, with almost no phase shift. The surface pressure variation is thus equal to $-\rho_0 V_{atm}$, where $\rho_0$ is the density of Titan's atmosphere at the surface. The deformation of the interior would therefore reduce the amplitude of the tidal pressure variations by $\gamma_2$, passing from $\sim$64 Pa at the sub-saturnian point, to likely less than 6.4 Pa for $\gamma_2<0.1$.
Fig. \ref{figure_3} shows the local amplitude of the tidal pressure variations ($\Delta P_{surf} = \gamma_2 \rho_0 V_{1}$) and phase compared to the sub-saturnian point ($\psi+\psi_2$), without interior deformation (left) and with interior deformation (right), using $1+\Re(k_2-h_2)$ = 0.08 and $\Im(k_2-h_2)$ = 0.01. These values correspond to an ice shell thickness of 80 km and a convective heat flux of 15 mW/m$^2$. We use these values as reference in the rest of the paper. In this case, $\gamma_2$ = 0.081, $\psi_2$ = -0.124 rad (corresponding to a time shift or -7.6 hours) and the amplitude of pressure variation at the sub-saturnian point is equal to $\Delta P_{surf}$ = 5.2 Pa.

\section{Computation of atmospheric tides}

\subsection{Analytical solution}
We derived the analytical solution of gravitational atmospheric tides caused by Saturn following the tidal theory from \cite{chapman1970} developed for the solar thermal tides and the lunar gravitational tides on Earth. This formalism was also used by \cite{strobel2006} for studying the development of gravitational tidal waves in Titan's upper atmosphere. In Titan's troposphere where winds are weak (geostrophic regime), gravitational atmospheric tides can be described by the following equations (the linearised primitive equations with no advective term) adding the gravitational tidal potential:
\begin{equation}
\boxed{
\begin{array}{rcl}
\frac{\partial u}{\partial t} - 2\Omega \sin(\lambda) v =-\frac{1}{R_{\rm T}\cos(\lambda)}\frac{\partial \left(\Phi  +V_{\rm atm} \right)} {\partial \phi} \\

\frac{\partial v}{\partial t} + 2\Omega \sin\lambda u =-\frac{1}{R_{\rm T}}\frac{\partial \left(\Phi  +V_{\rm atm} \right)} {\partial \lambda} \\

\frac{1}{R_{\rm T} \cos\lambda} \left(  \frac{\partial u}{\partial \phi} +   \frac{\partial v \cos\lambda}{\partial \lambda} \right) + \frac{1}{\rho_0}\frac{\partial \rho_0 w}{\partial z} = 0 \\

\frac{\partial^2 \Phi}{\partial z \partial t } + N^2w = 0
\end{array}
}
\label{eq8}
\end{equation}
where $u$, $v$, and $w$ are the zonal, meridional and vertical winds, $\rho_0$ is the air density, $\Phi$ is the geopotential, $p$ is the pressure. $N=\sqrt{\frac{g}{\theta}\frac{d\theta}{dz}}=\sqrt{\frac{g}{T}(\Gamma_d-\Gamma)}$ is the Brunt-V\"ais\"al\"a frequency ($\theta$ is the potential temperature,  $\Gamma$ and $\Gamma_d$ are the temperature gradient and the adiabatic temperature gradient of the atmosphere). $N\sim$ 0.001 s$^{-1}$ in the first 2 km of Titan's troposphere \citep{charnay2012}.
The atmosphere is forced by a tidal potential periodic in longitude and time. We search complex solutions of the form: 
\begin{equation}
(u, v, w, \Phi, V_{\rm atm}) = Re[(\hat{u}, \hat{v}, \hat{w}, \hat{\Phi}, \hat{V}_{\rm atm})e^{i(s\phi+2\Omega\sigma t)}]
\end{equation}
with $\sigma$=1 and $s$=-1/2 for the eastward mode, $\sigma$=1 and $s$=1/2 for the westward mode, and $\sigma$=1 and $s$=0 for the stationary mode. Solutions satisfy the equations:
\begin{equation}
\boxed{
\begin{array}{rcl}
\hat{u}=\frac{-\sigma}{2a\Omega}\mathcal{S}^{\sigma,s}_u(\hat{\Phi}+\hat{V}_{atm}) \\

\hat{v}=\frac{i\sigma}{2a\Omega}\mathcal{S}^{\sigma,s}_v(\hat{\Phi}+\hat{V}_{atm}) \\

\mathcal{L}^{\sigma,s}\hat{\Phi} - i \frac{2R_{\rm T}^2 \Omega}{\sigma \rho_0}\frac{\partial(\rho_0 \hat{w})}{\partial z} = -\mathcal{L}^{\sigma,s}\hat{V}_{atm}  \\

2i \Omega \sigma \frac{\partial\hat{\Phi}}{\partial z} + N^2 \hat{w} = 0
\end{array}
}
\end{equation}
where $\mathcal{L}^{\sigma,s}$ is the Laplace tidal operator given by:
\begin{equation}
\mathcal{L}^{\sigma,s}\hat{\Phi} = \partial_\mu \left(\frac{1-\mu^2}{\sigma^2-\mu^2}\partial_\mu\hat{\Phi} \right) - \frac{1}{\sigma^2-\mu^2} \left(\frac{s(\sigma^2+\mu^2)}{\sigma(\sigma^2-\mu^2)} + \frac{s^2}{1-\mu^2} \right) \hat{\Phi}
\end{equation}
$\mathcal{S}^{\sigma,s}_u$ and $\mathcal{S}^{\sigma,s}_u$ are the horizontal wind operators given by:
\begin{equation}
\begin{split} 
& \mathcal{S}^{\sigma,s}_u =  \left[\frac{s}{\sigma^2-\mu^2} - \frac{\mu(1-\mu^2)}{\sigma(\sigma^2-\mu^2)}\frac{\partial}{\partial\mu} \right] \\
& \mathcal{S}^{\sigma,s}_v =  \left[\frac{s}{\sigma(\sigma^2-\mu^2)} - \frac{(1-\mu^2)}{(\sigma^2-\mu^2)}\frac{\partial}{\partial\mu} \right]
\end{split} 
\end{equation}
with $\mu=\sin\lambda$. Eigenvectors of the Laplace tidal operator are the Hough functions $\Theta_n^{\sigma,s}$ which verify:
\begin{equation}
\mathcal{L}^{\sigma,s} \Theta_n^{\sigma,s} + \gamma_n^{\sigma,s} \Theta_n^{\sigma,s} = 0
\end{equation}
The eigenvalues $\gamma_n^{\sigma,s}$ are called the Lamb parameters, also defined with equivalent heights $h_n^{\sigma,s}$ with $\gamma_n^{\sigma,s} = 4\Omega^2R_{\rm T}^2/(g h_n^{\sigma,s})$

If we decompose the tidal potential and solutions in terms of Hough functions:  $\hat{V}_{\rm atm}=\sum\limits_{n=1}^{\infty}V^{\sigma,s}_n \Theta_n^{\sigma,s}$ and $(\hat{u}, \hat{v}, \hat{w}, \hat{\Phi})=\sum\limits_{n=1}^{\infty}\left(\hat{u}_n, \hat{v}_n, \hat{w}_n, \hat{\Phi}_n \right)\Theta_n^{\sigma,s}e^{z/2H}$, the problem is reduced to the resolution of an equation on the vertical structure:
\begin{equation}
\frac{d^2\hat{w}_n}{dz^2} +\left(\frac{N^2}{gh_n^{\sigma,s}} -  \frac{1}{4H^2}     \right)\hat{w}_n = 0
\end{equation}
with as boundary condition at the surface $\frac{D\Phi}{Dt}=0$ which can be expressed as:
\begin{equation}
\frac{d\hat{w}_n}{dz} +\left(\frac{1}{h_n^{\sigma,s}} -  \frac{1}{2H}     \right)\hat{w}_n = \frac{2i\sigma\Omega V^{\sigma,s}_n}{gh_n^{\sigma,s}}
\end{equation}

\begin{figure}[h!] 
\begin{center} 
\includegraphics[width=8cm,clip]{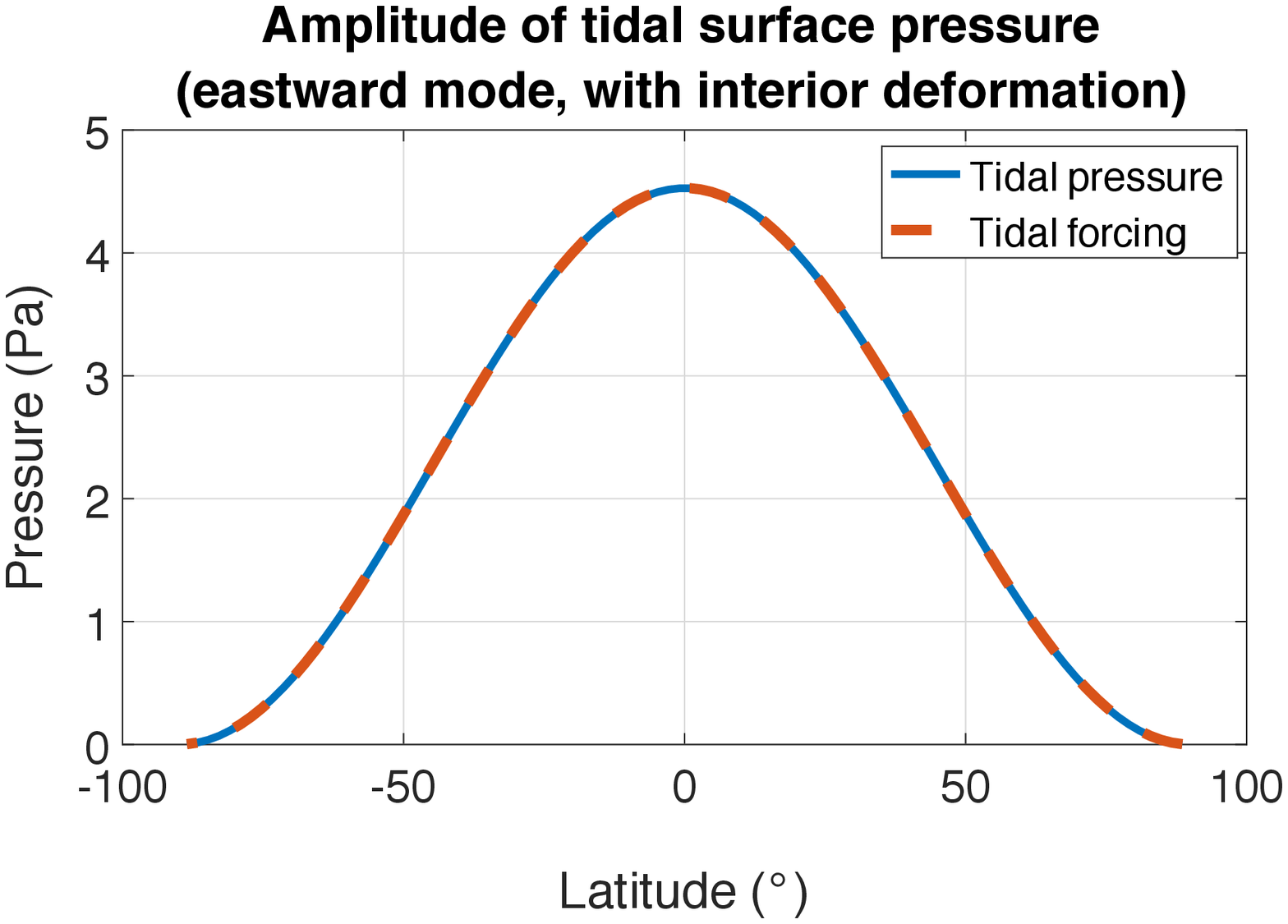}
\includegraphics[width=8cm,clip]{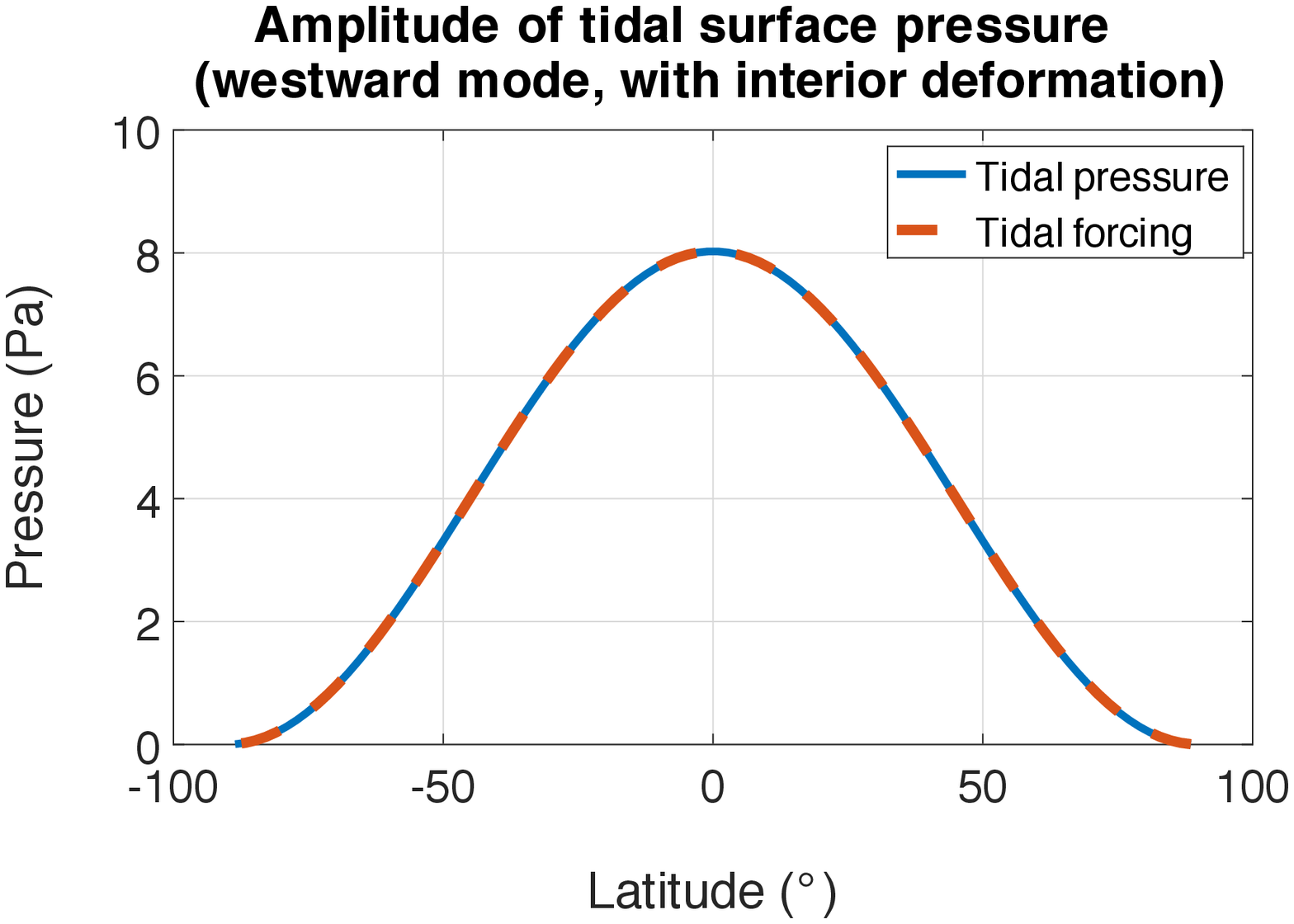}
\end{center} 
\caption{Amplitude of the atmospheric surface pressure for the eastward (top) and westward (left) tidal wave compared to the tidal forcing. Analytical calculation including interior deformation with $1+\Re(k_2-h_2$)=0.08 and $\Im(k_2-h_2$)=0.01.}
\label{figure_4}
\end{figure}

\begin{figure*}[h!] 
\begin{center} 
\includegraphics[width=8cm,clip]{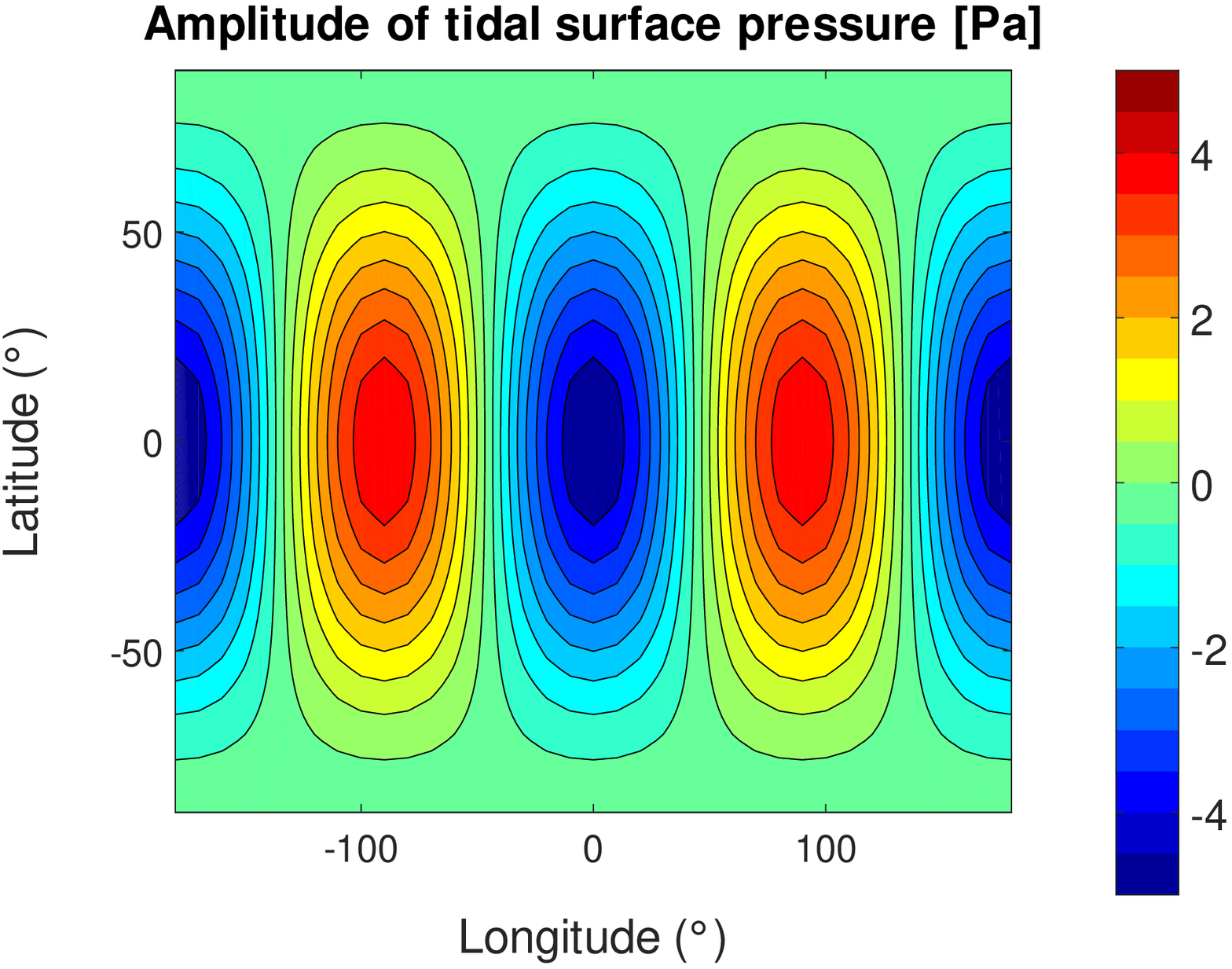}
\includegraphics[width=8cm,clip]{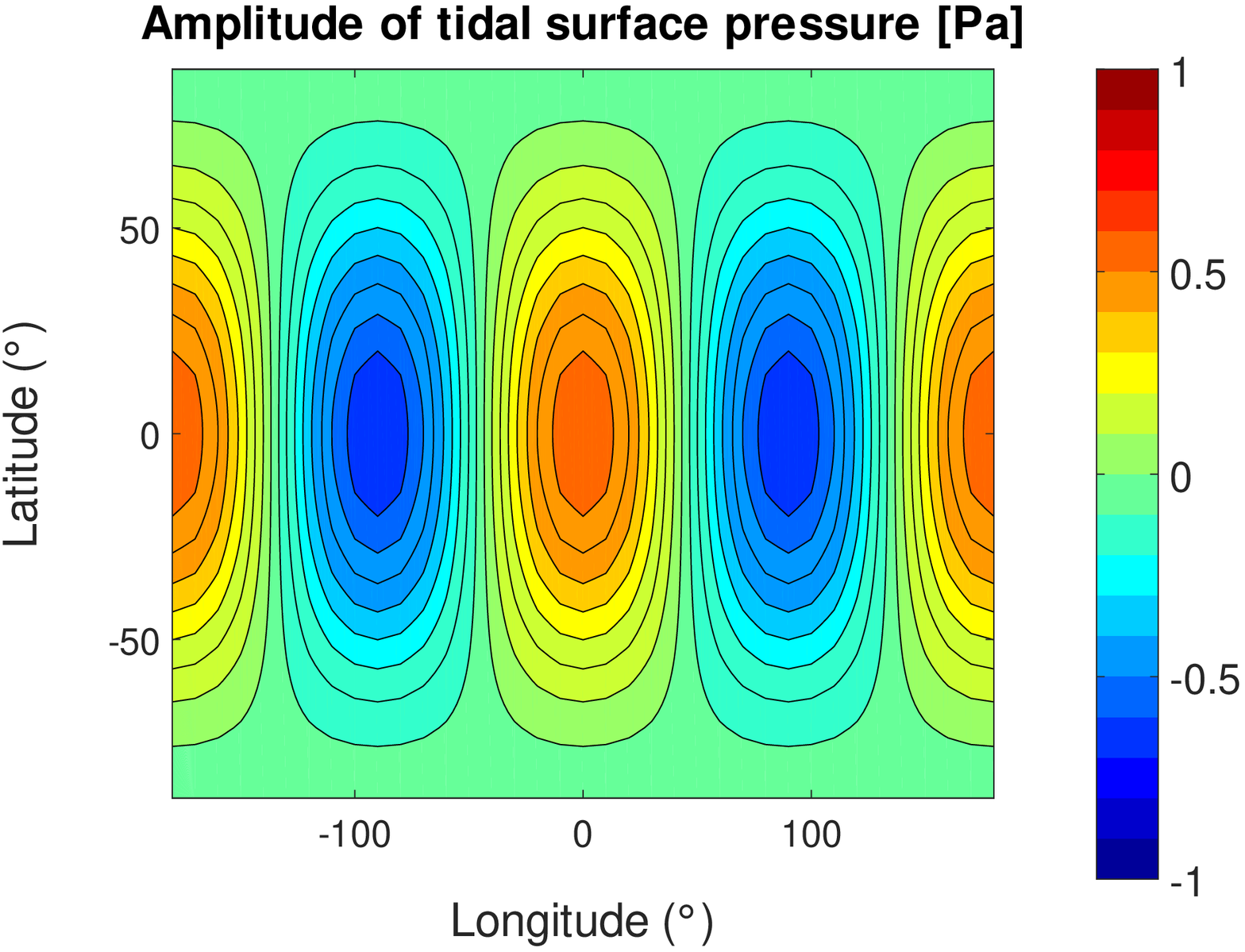} \\
\includegraphics[width=8cm,clip]{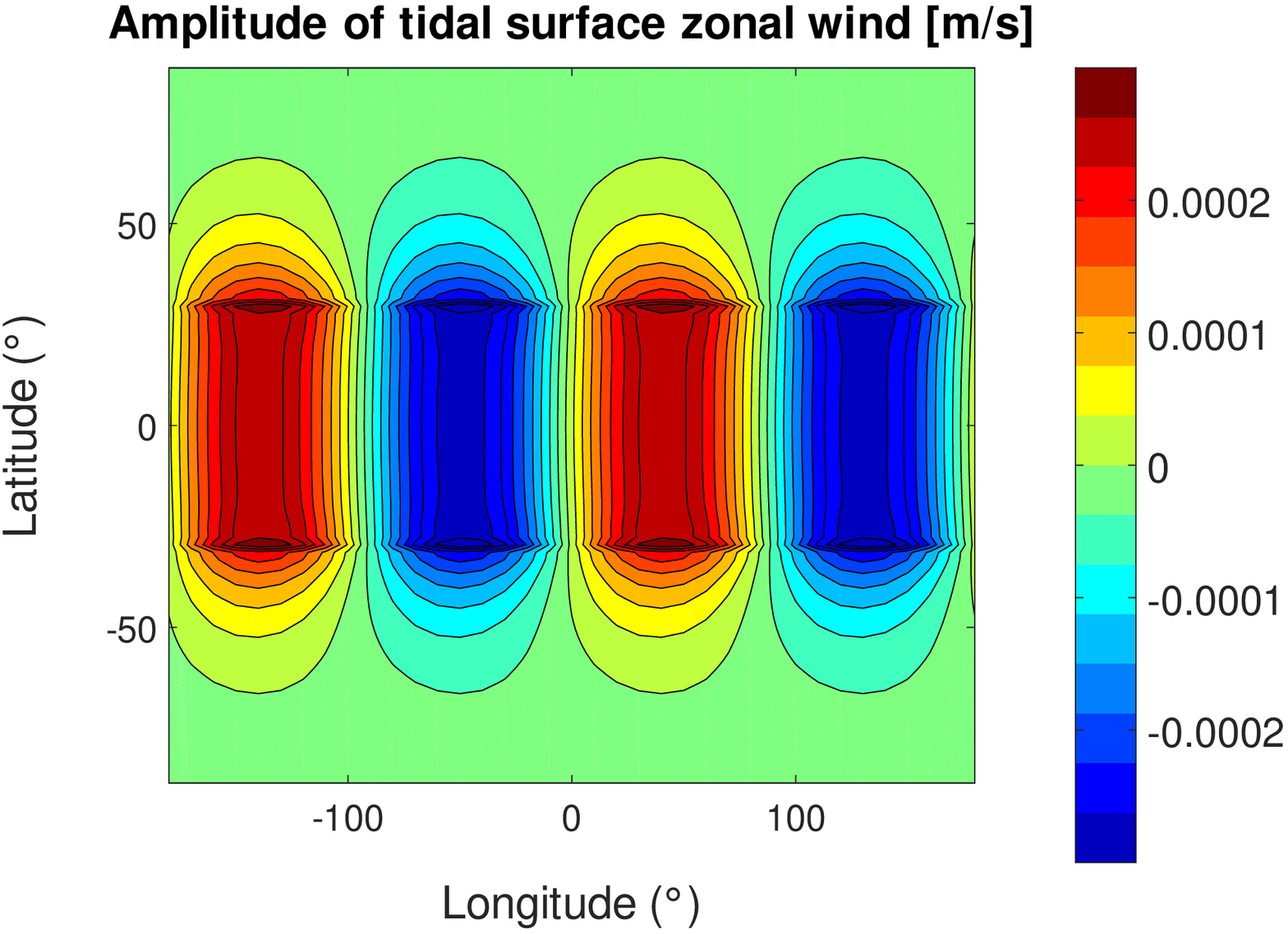}
\includegraphics[width=8cm,clip]{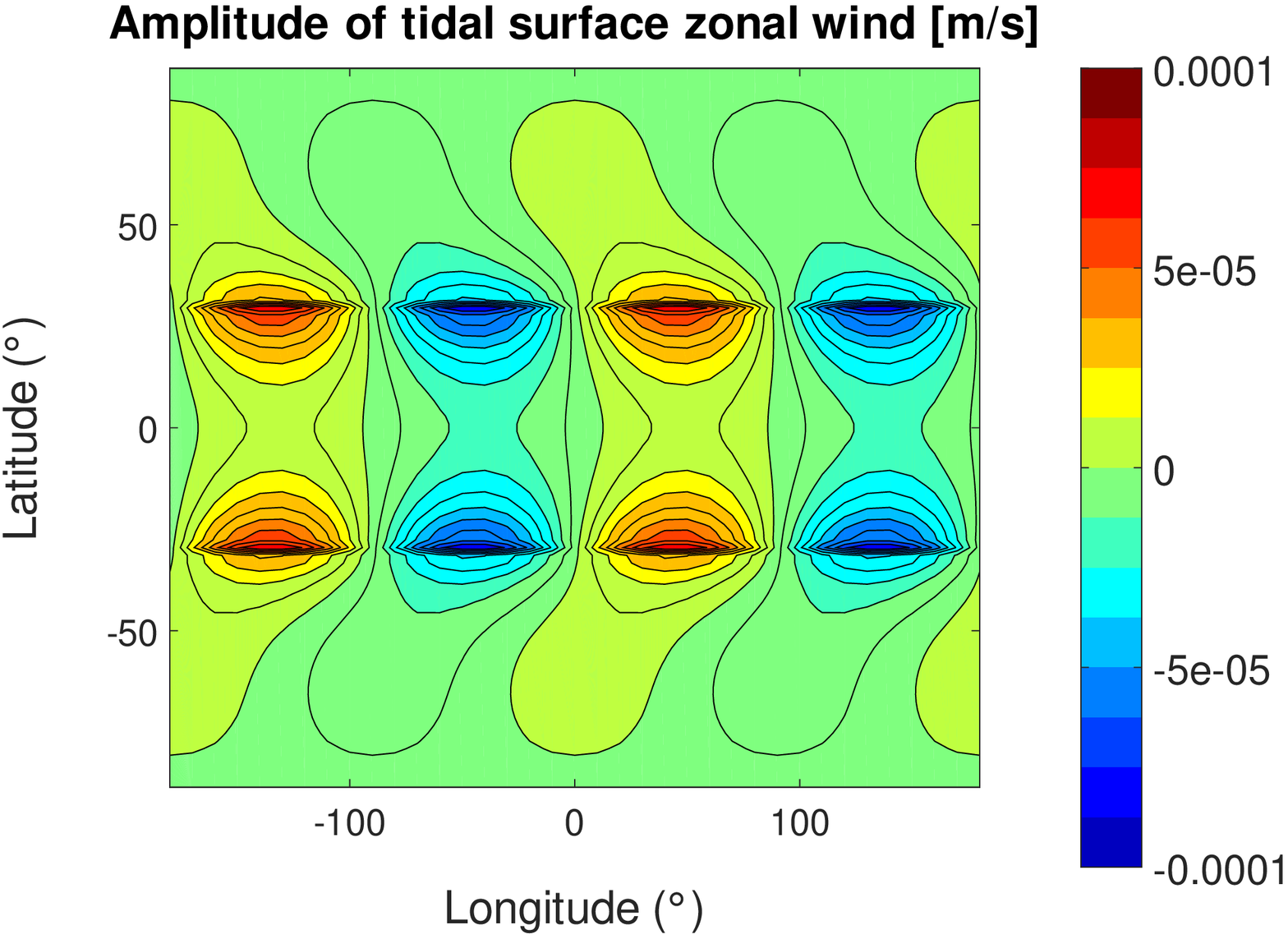} \\
\includegraphics[width=8cm,clip]{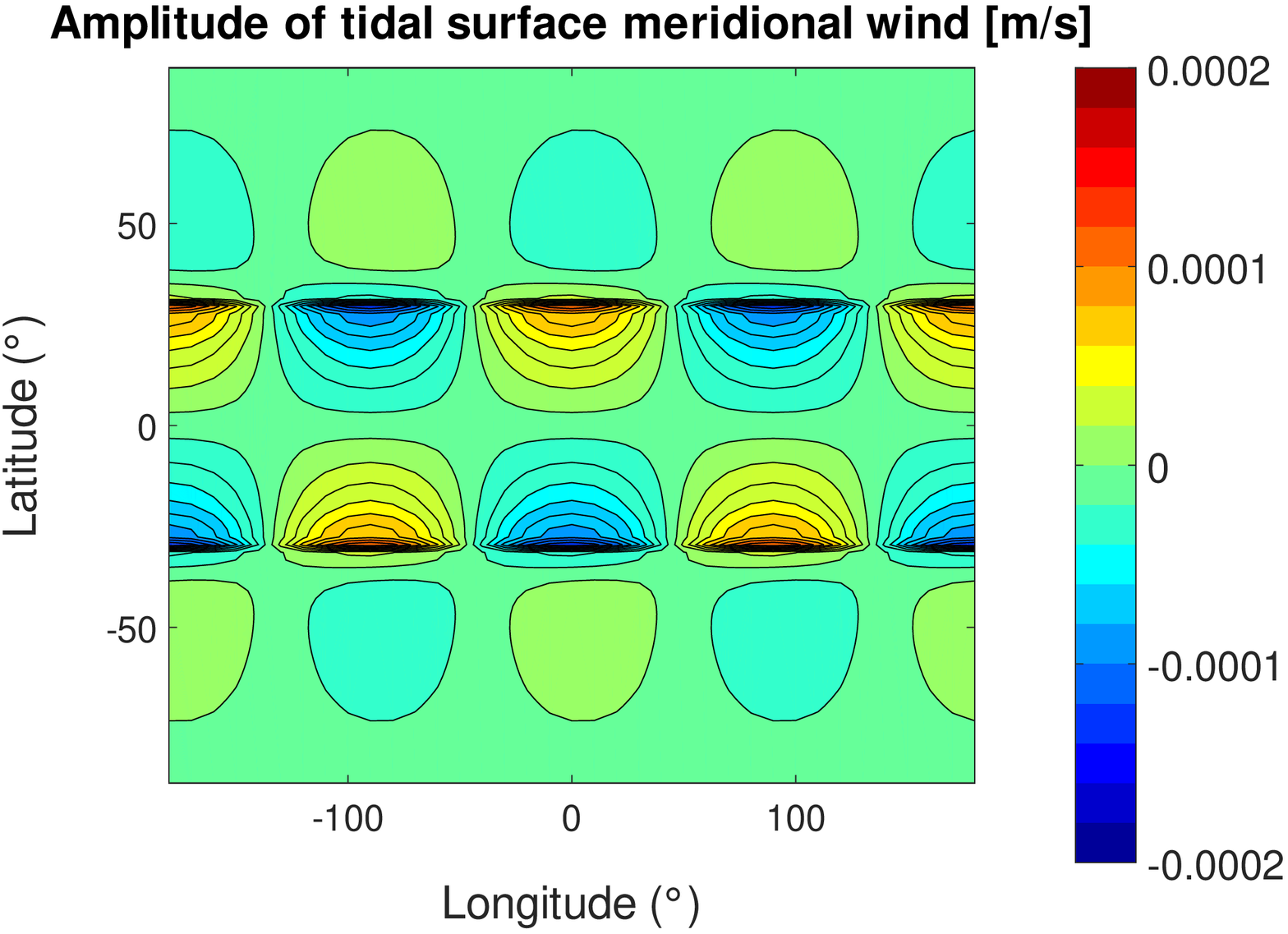}
\includegraphics[width=8cm,clip]{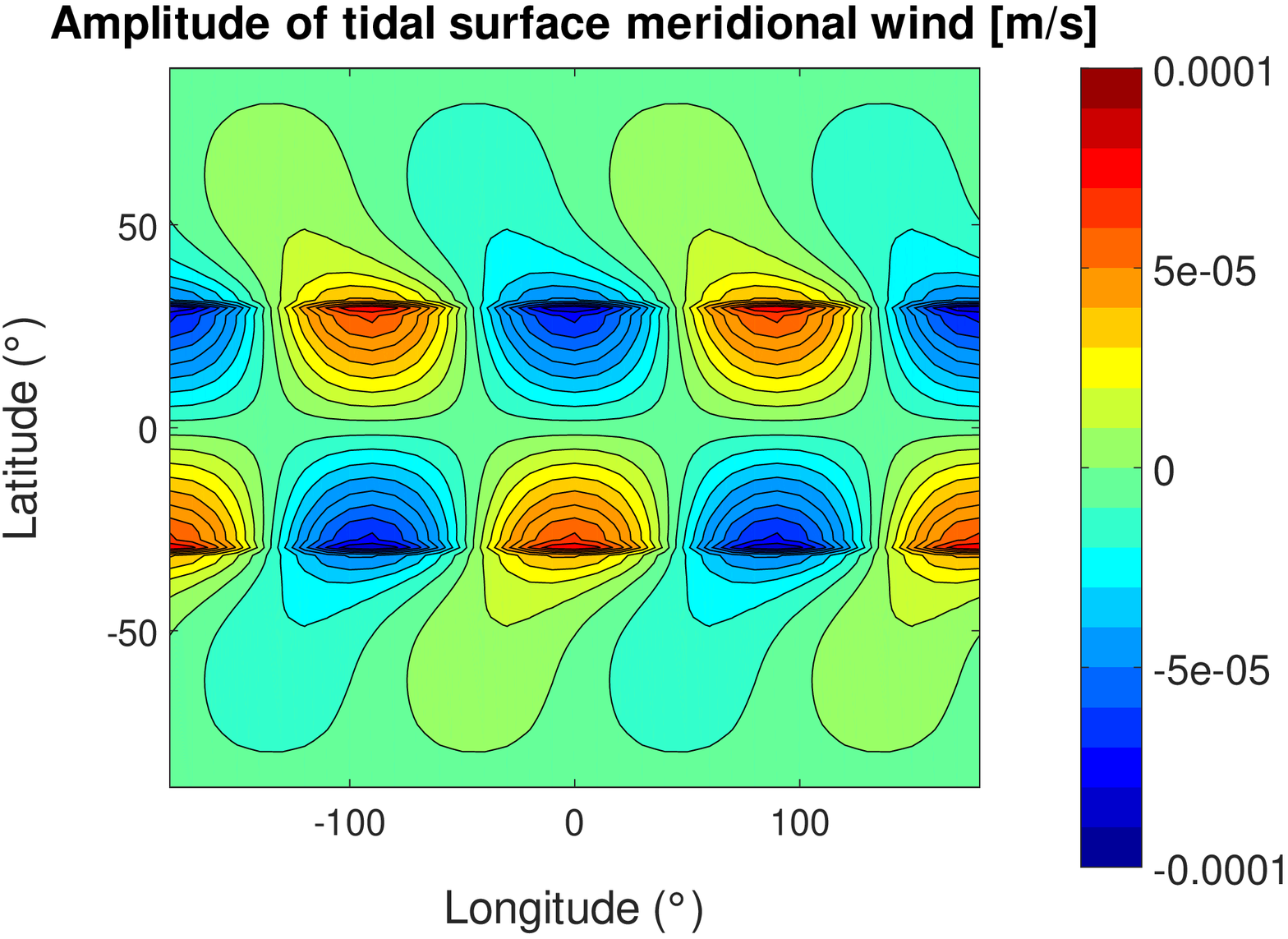} \\
\includegraphics[width=8cm,clip]{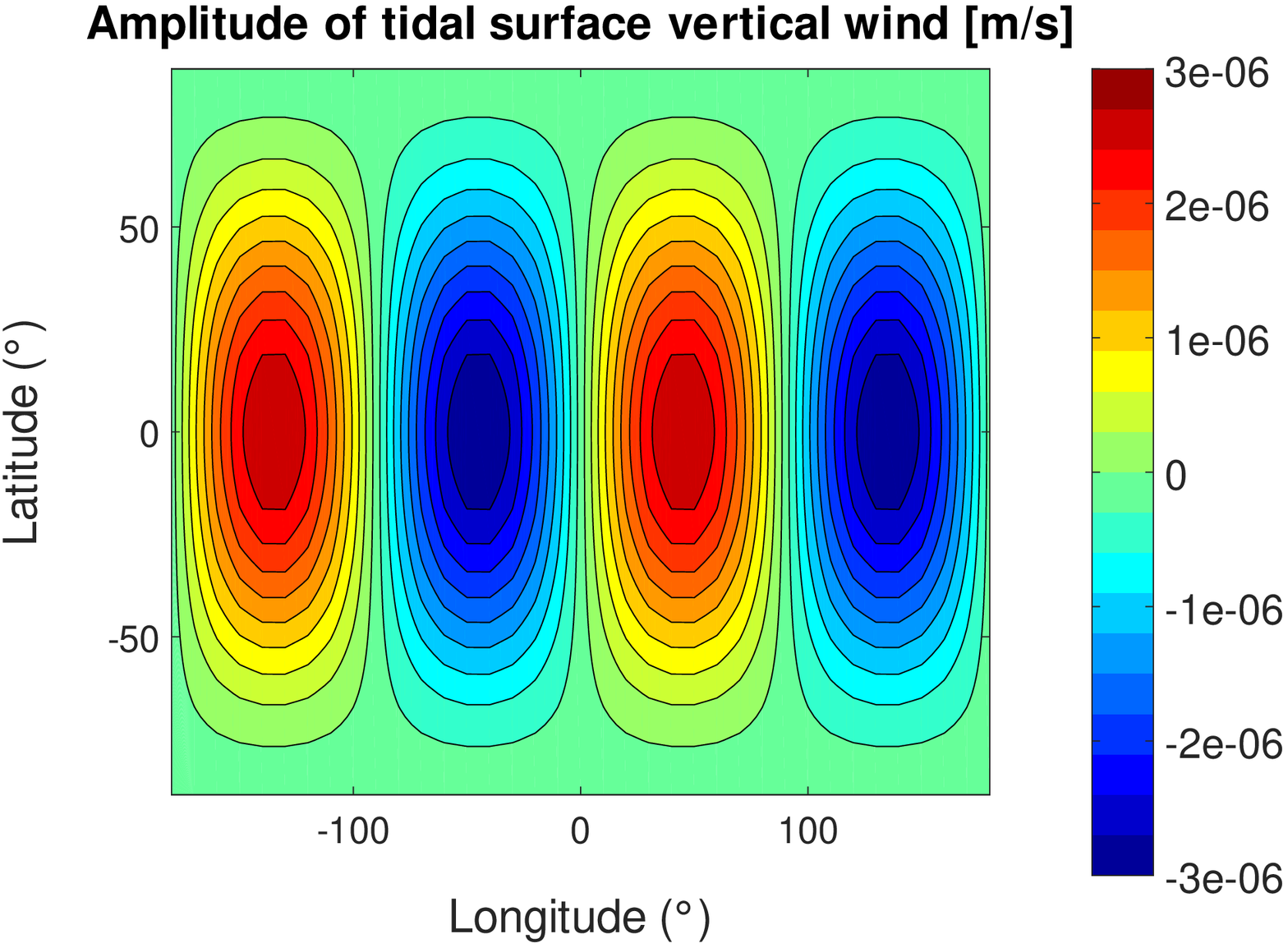}
\includegraphics[width=8cm,clip]{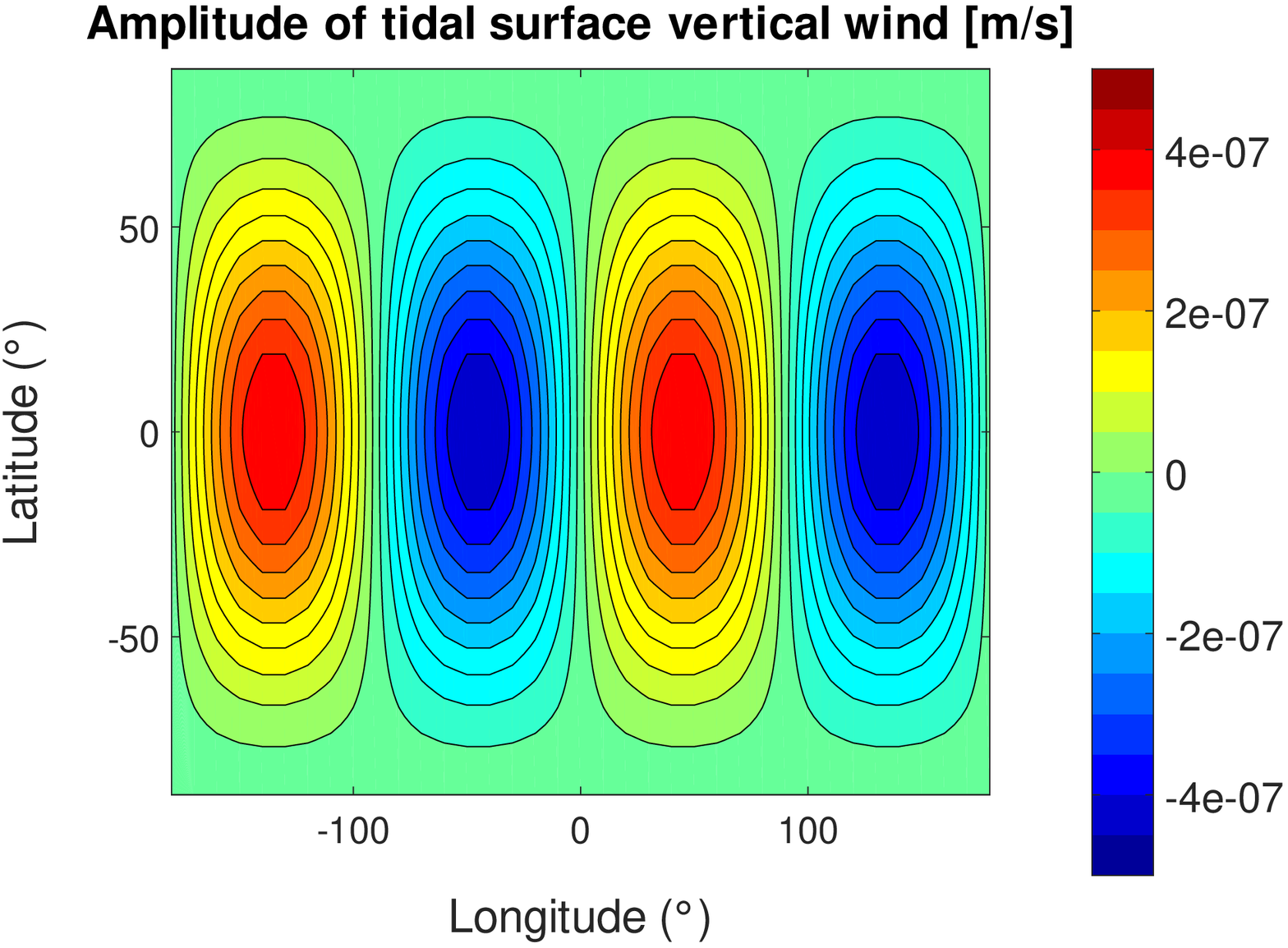} 
\end{center} 
\caption{Tidal surface pressure, and tidal zonal, meridional and vertical surface winds as a function of longitude and latitude at t=0 for the eastward wave (left) and the westward wave (right).
Analytical calculation including interior deformation with $1+\Re(k_2-h_2$)=0.08 and $\Im(k_2-h_2$)=0.01.}
\label{figure_5}
\end{figure*}

To compute analytically the pressure and wind variations, we first computed the Hough functions and Lamb parameters for given values of $\sigma$ and $s$, using the algorithm from \cite{wang2016}. Then, we expanded the tidal potential in Hough functions. We solved the vertical structure equation for each mode n, assuming an upward wave propagation for $(\frac{1}{h_n^{\sigma,s}}-\frac{1}{2H}) \ge 0$, and an evanescent wave for $(\frac{1}{h_n^{\sigma,s}}-\frac{1}{2H})<0$. Since $h_n^{\sigma,s}<<H$ for all modes, the wave propagation only depends on the sign of $h_n^{\sigma,s}$. Then we solved the surface boundary condition and  we determined $\hat{u}_n$, $\hat{v}_n$, $\hat{w}_n$ and $\hat{\phi}_n$ in complex form. Finally we summed results over n to get the pressure and wind fields. Fig. \ref{figure_4} shows the amplitude of the surface pressure variation for the eastward and the westward tidal waves. The surface pressure perfectly follows the tidal forcing for both cases.
Fig. \ref{figure_5} shows maps of tidal surface pressure and winds at a given time for the eastward and westward modes. The maximal tidal wind speed is 3$\times 10^{-4}$ m/s, which is three orders of magnitude lower than the typical surface wind speed predicted by 3D GCMs (e.g. \cite{lebonnois2012,charnay2012}).

\subsection{Simulations with the Titan IPSL GCM}

We performed 3D simulation with the Titan IPSL GCM \citep{lebonnois2012} including the tidal potential caused by Saturn. We used a horizontal
resolution of 32$\times$48 (corresponding to resolutions of 11.25$^\circ$ longitude by 3.75$^\circ$ latitude) and 55 vertical levels.
Simulations were run for 10 Titan days at the vernal equinox with thermal tides and with/without gravitational tides. Dragonfly will operate during the southern summer, but we did not notice influence of the season on the amplitude of gravitational tides. We used the tidal potential without or with interior deformation (with $1+\Re(k_2-h_2)$ = 0.08 and $\Im(k_2-h_2)$ = 0.01). We analysed simulations from day 5 to day 10, removing the first 5 days during which the atmosphere readjusts to the presence/absence of tides. 

\begin{figure}[h!] 
\begin{center} 
\includegraphics[width=8cm,clip]{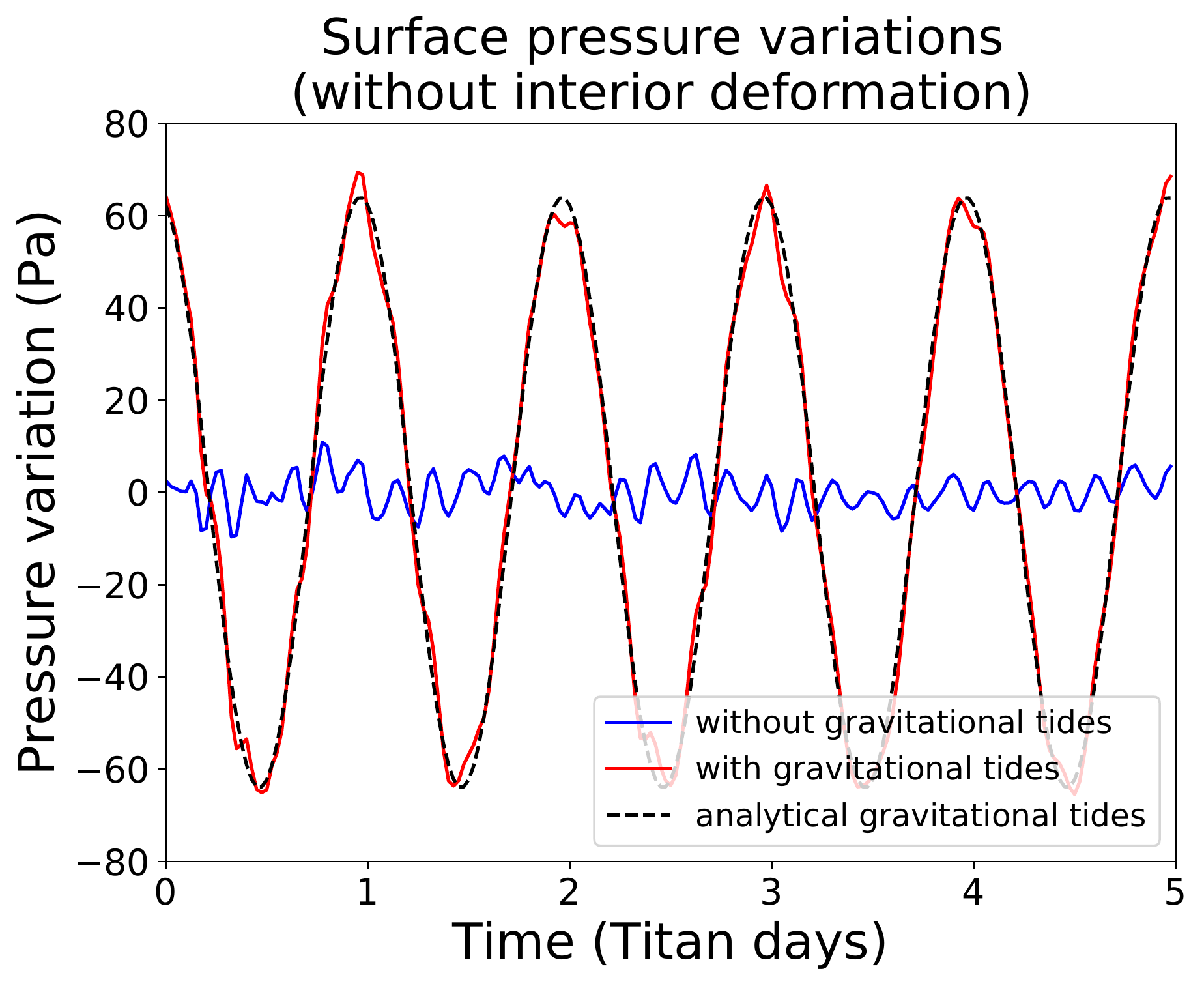}
\includegraphics[width=8cm,clip]{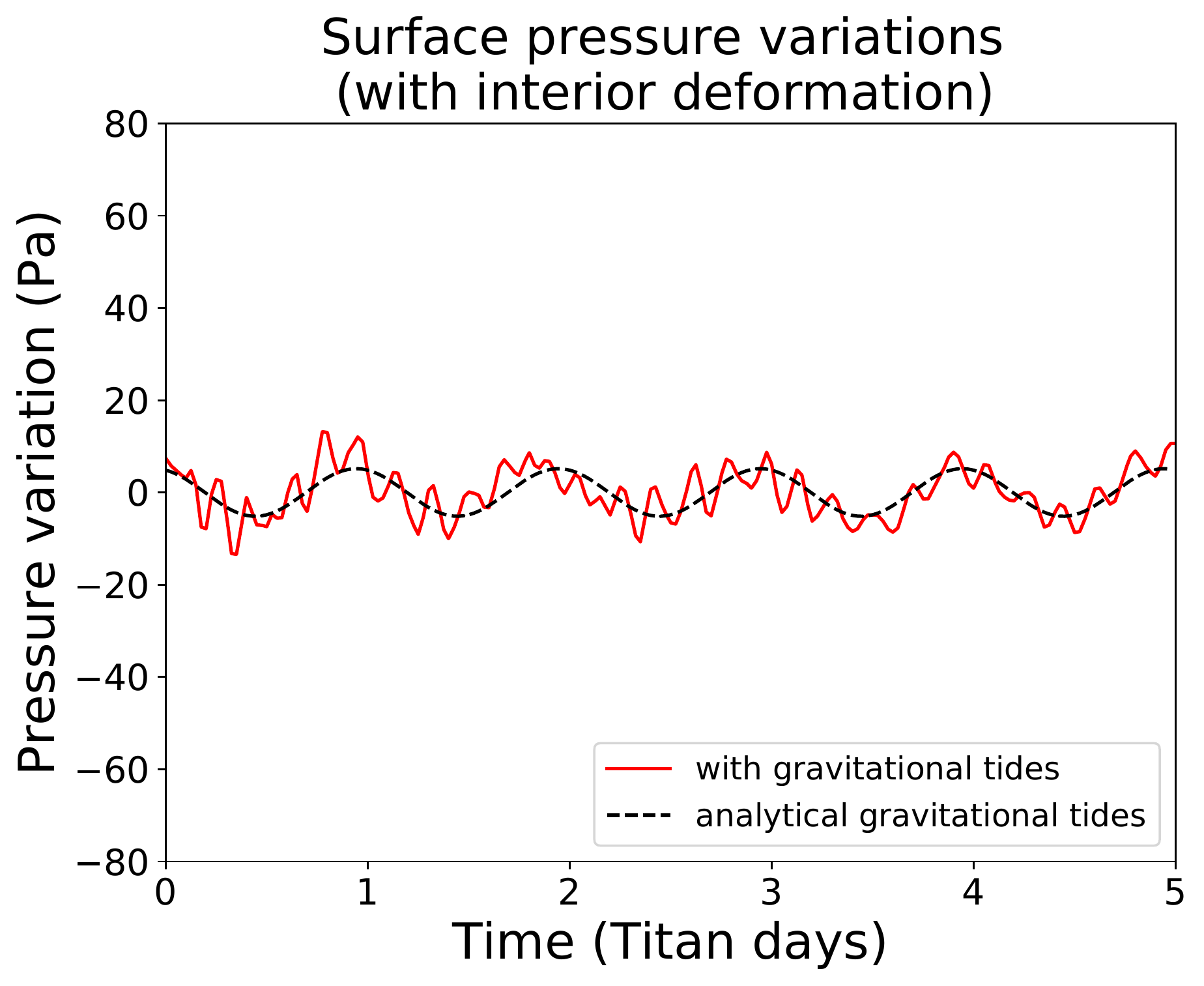}
\end{center} 
\caption{Pressure variations at longitude 0$^\circ$ and latitude 0$^\circ$N computed with the Titan IPSL GCM. The top (bottom) panel shows the effects of gravitational tides without (with) interior deformation. The case with interior deformation assumed $1+\Re(k_2-h_2$)=0.08 and $\Im(k_2-h_2$)=0.01.
Thermal tides were included in all simulations. The blue line is the top panel shows the pressure variations without gravitational tides for comparison. The analytical tidal pressure variation is shown with a black dashed line.}
\label{figure_6}
\end{figure}

Simulated pressure and horizontal wind variations confirm the analytical calculations from the previous subsection. The pressure field follows the tidal potential, and tidal winds are very weak in the simulations. 
Fig. \ref{figure_6} shows the time-evolution of the surface pressure at the sub-saturnian point (longitude 0$^\circ$ and latitude 0$^\circ$) with thermal tides (diurnal solar cycle) and with/without gravitational tides. For the case without interior deformation (left panel), the gravitational tides dominate the surface pressure variations which follow well the tidal potential. 
The standard deviation of the surface pressure without gravitational tides or substracting the theoretical tidal pressure is around 4 Pa. This is less than the amplitude of the tidal pressure variations for the case with interior deformation (i.e. $\Delta$P = 5.2 Pa). Although small, the tidal signal is clearly identified in the simulation with interior deformation (right panel).

\begin{figure}[h!] 
\begin{center} 
\includegraphics[width=8cm,clip]{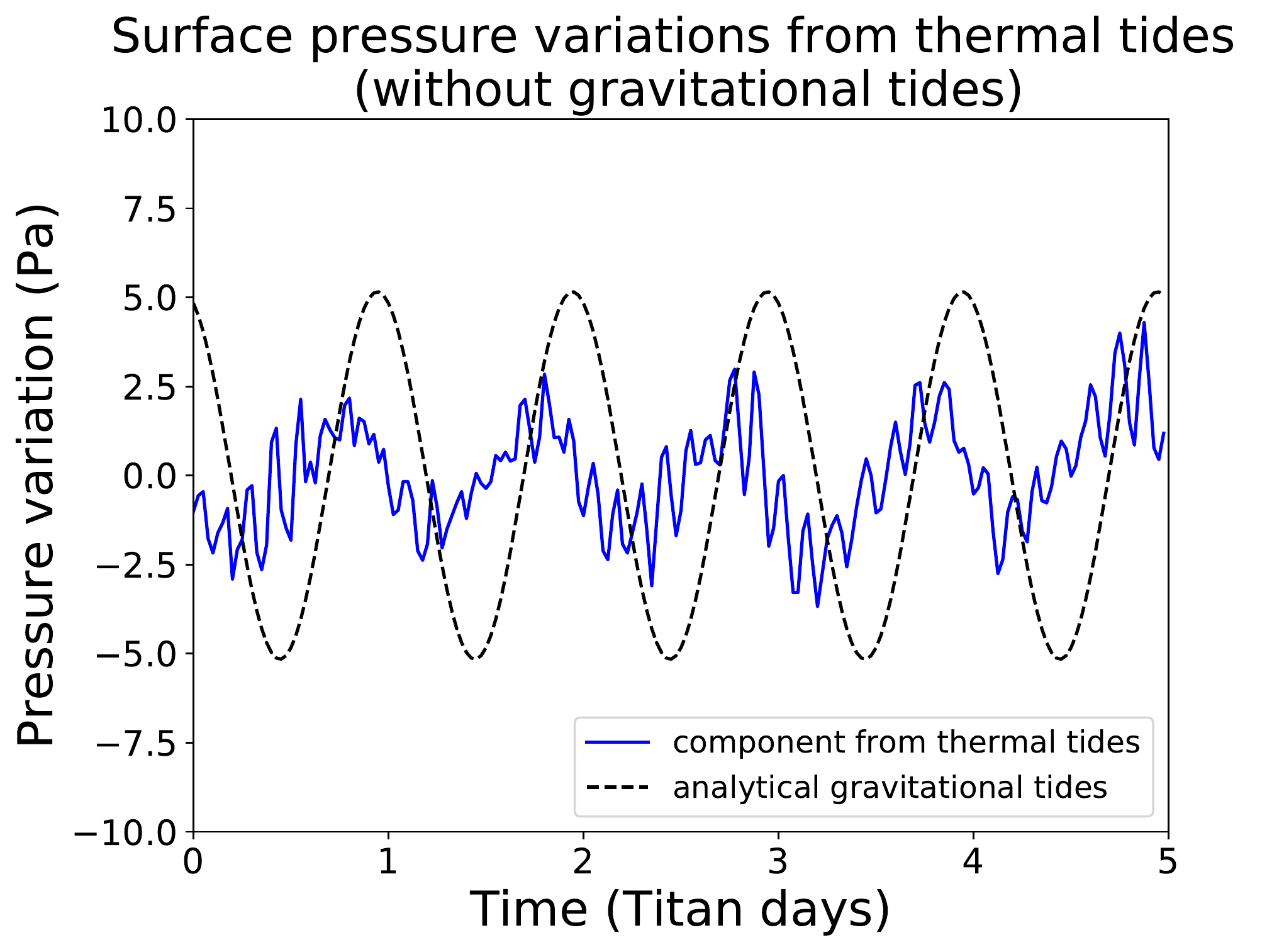} 
\end{center} 
\caption{Pressure variations caused by thermal tides at longitude 0$^\circ$ and latitude 0$^\circ$N computed with the Titan IPSL GCM. The analytical pressure variation from gravitational tides with interior deformation is shown with a black dashed line.}
\label{figure_7}
\end{figure}

We found that the gravitational tides have a negligible effect on the surface temperature (variations of $\sim 10^{-3}$ K correlated with surface pressure for the case with  deformation of the interior), which is dominated by the diurnal solar cycle (peak-to-peak variations of $\sim$0.5 K). Fig. \ref{figure_7} shows the surface pressure variations from thermal tides for the case without gravitational tides.
It was computed by making the direct substraction of a simulation with thermal tides minus without thermal tides.
It reveals that the surface pressure variations from thermal tides are dominated by the diurnal mode, as also found by \cite{tokano2010}. It has an amplitude of 1.7 Pa, smaller than gravitational tides and with a phase shift (the maximum occurs at $\sim$6 pm solar local time).
In contrast, the pressure variation from thermal tides on Earth are dominated by the semi-diurnal mode \citep{chapman1970}. It is forced in particular by the solar energy absorption by stratospheric ozone and has two pressure maximums at $\sim$10 a.m. and $\sim$10 p.m. solar local time.

In our simulations, a major source of variability in the lower troposphere comes from baroclinic waves, which develop at mid-latitudes with a period of 2-5 Titan days and a pressure variation of $\sim$20 Pa \citep{lebonnois2012}. However, their impact is limited in the equatorial region (pressure variation of $\sim$1 Pa) and they can be distinguished from gravitational tides by their long period. \cite{mitchell2011} also found that equatorially trapped Kelvin waves with a period of $\sim$0.5 Titan day would occur. They are likely responsible for the large chevron-shaped methane storm observed at the equator during the equinoctial season \citep{turtle2011}. We did not identify these modes in our simulations. The variability caused by planetary waves in the equatorial region may thus be higher than predicted by our model, in particular close to the equinox when equatorial clouds and storms form.

\begin{figure}[h!] 
\begin{center} 
\includegraphics[width=8cm,clip]{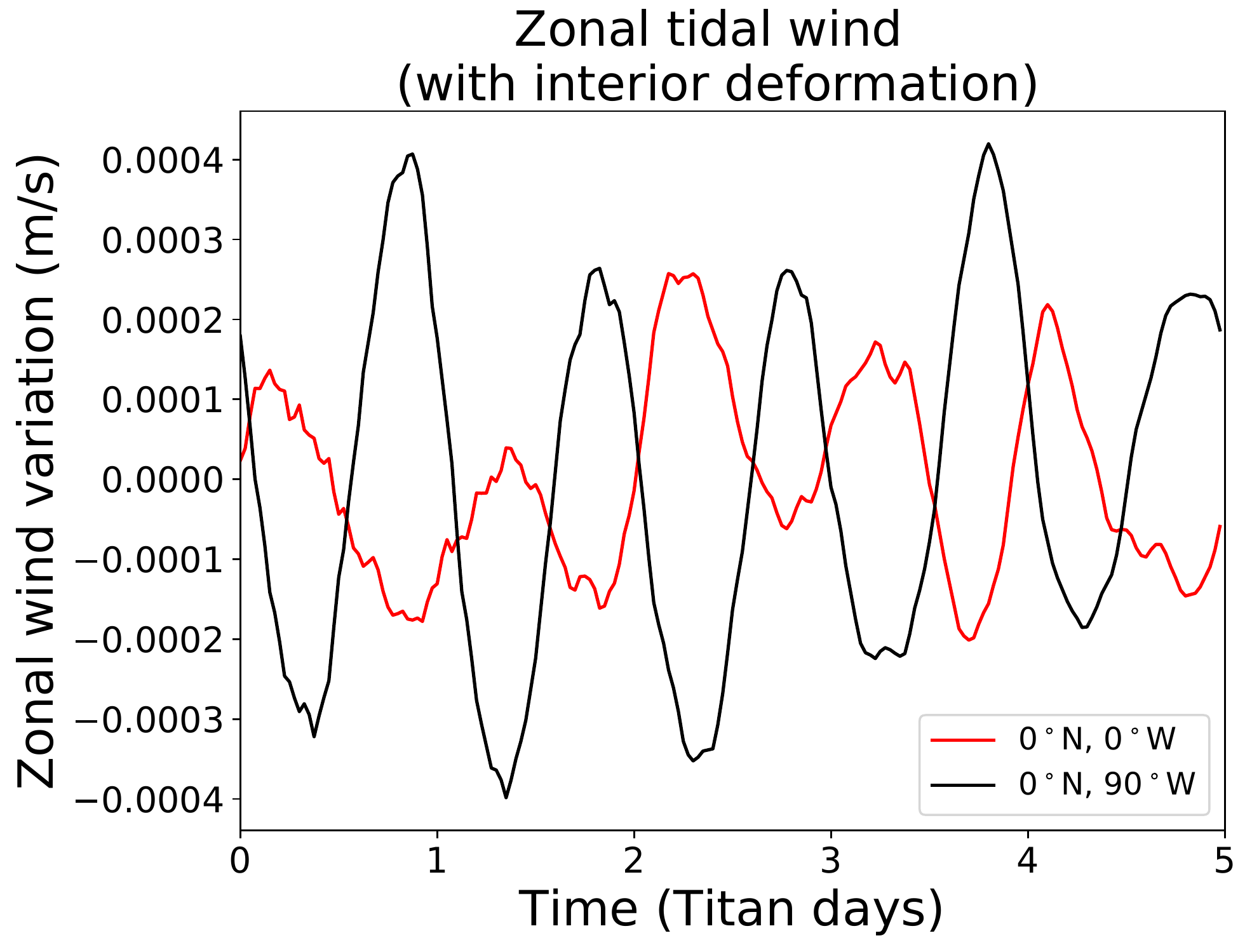}
\end{center} 
\caption{Component of zonal tidal winds at 35 m above the surface computed at latitude 0$^\circ$N, and longitude 0$^\circ$W (red) or 90$^\circ$W (black). The lines corresponds to the difference between simulations with tides and without tides. Thermal tides were not included in these simulations.}
\label{figure_8}
\end{figure}

Fig. \ref{figure_8} shows the zonal tidal wind component at 35 m above the surface for the case with interior deformation. It was computed without thermal tides to eliminate their effect and by making the direct substraction of a simulation with gravitational tides minus without gravitational tides. From day 5 to day 10, the wind field is sufficiently close between the two simulations (with/without gravitational tides) that the difference of surface wind corresponds quite well to the tidal component. The simulated tidal wind component is extremely weak, with an amplitude of $\sim 3 \times 10^{-4}$ m/s at the sub-saturnian point. This is much lower than typical surface winds which are around 0.5 m/s in the IPSL GCM \citep{lebonnois2012, charnay2012}. This tidal wind speed is consistent with our analytical calculations (see Section 2.2). \cite{friedson2009} performed 3D simulations of Titan's atmosphere including gravitational tides with the Titan-CAM model. They found an amplitude of tidal winds of around 0.05 m/s compared to 0.5 m/s in the simulations from \cite{tokano2002}. Our simulated tidal winds are respectively two orders of magnitudes and three orders of magnitudes lower than those in the simulations from \cite{friedson2009} and \cite{tokano2002}. The deformation of the interior is responsible for a reduction of tidal wind speed by one order of magnitude, meaning that large discrepancies remain between the tidal responses of GCMs for the same forcing. These discrepancies could be related to the mean circulation or to the thermal structure.
In particular, the tidal wind amplitude is proportional to $N^2$ (the squared Brunt-V\"ais\"al\"a frequency) and thus very sensitive to the thermal structure. The lower troposphere may be more adiabatic in the IPSL GCM than in the models by \cite{tokano2002} and \cite{friedson2009}, leading to weaker tidal winds. In this regard, the Titan IPSL GCM reproduces well the thermal structure and the zonal wind measured by the Huygens probe in Titan's troposphere, especially in the planetary boundary layer \citep{lebonnois2012, charnay2012}, giving confidence in our predictions. According to our analytical and numerical calculations, tidal winds should not be detectable in Titan's lower troposphere.

\subsection{Impact of tides on cloud formation}
The tidal pressure variation and the associated adiabatic heating/warming induces a change of relative humidity, which can impact cloud formation and precipitation. A statistical analysis of 15 years of data from ERA-Interim shows that the lunar semidiurnal tide induces a variation of relative humidity and precipitation with an amplitude of 0.039$\%$ and 0.4$\%$ respectively on Earth \citep{kohyama2016}. 
One may wonder if the stronger saturnian gravitational tides could affect cloud formation on Titan, with a preferential formation where the amplitude of tides are maximal (i.e. close to the subsaturnian and anti-saturnian point). This mechanism has been suggested to explain the two maximums of cloud occurrence from Cassini close to longitudes 0$^\circ$E and 180$^\circ$E \citep{rodriguez2009}.

The passage of the tidal wave is associated with an adiabatic warming/cooling :
\begin{equation}
dT=-\frac{R^*T}{c_p P}dP
\end{equation}
with $R^*$ the specific gas constant, $c_p$ the specific heat capacity of Titan's atmosphere, T the temperature and
P the pressure. Above the surface, the amplitude of temperature variations caused by tides is $\sim 1 \times 10^{-3}$ K.
The relative humidity of methane clouds defined as $RH=\frac{P_{vap}}{P_{sat}}$ evolves as:
\begin{equation}
\frac{dRH}{RH}=\frac{dP_{vap}}{P_{vap}}-\frac{dP_{sat}}{P_{sat}}
\end{equation}

If we assume no condensation, $\frac{dP_{vap}}{P_{vap}}=\frac{dP}{P}$. The Clausius-Clayperon relation gives $\frac{dP_{sat}}{dT}=\frac{L_{CH_4}P_{sat}}{R_{CH_4}T^2}$, where $L_{CH_4}$ is the mass-latent heat of methane and $R_{CH_4}$ is the specific gas constant of methane. These relations lead to:
\begin{equation}
\frac{dRH}{RH}=\frac{dP}{P}\left(1-\frac{L_{CH_4}R^*}{C_p R_{CH_4}T} \right)
\end{equation}

For Titan's tropospheric conditions, $\frac{L_{CH_4}R^*}{C_p R_{CH_4}T} \sim 3.1$. The relative humidity is maximal when the tidal pressure is minimal. The maximal amplitude of relative humidity variation is around 0.01$\%$. The effect is therefore very small, even four times smaller than the lunar tides on the Earth. We conclude that the influence of gravitational tides on Titan's clouds should not be detectable in Cassini data. Longitudinal variations of the cloud occurrence rate should rather be due to variations of surface properties (topography, surface humidity, albedo or thermal inertia).

\section{Implications for the Dragonfly mission}

\subsection{Dragonfly's ability to measure atmospheric tides}

The Dragonfly mission is discussed in \cite{turtle2018},  \cite{barnes2021} and \cite{lorenz2018}. In addition to the pressure/tide determination approach in the present paper, Dragonfly may also constrain the thickness of Titan's ice shell via the Schumann resonance \citep{lorenz2020}, and with seismic methods \citep{lorenz2019}. Note that Dragonfly (now planned to launch in 2027) is still in an early design phase, and some of the following details may evolve. 
The Dragonfly initial landing site \citep{lorenz2021} is just south of the Selk crater, at 7$^\circ$ N 199$^\circ$W. This is fortunately rather close to the anti-Saturnian point, where the pressure tide amplitude is nearly at its maximum value (see Fig. \ref{figure_1}). 
 
The pressure measurement on Dragonfly's DraGMet (Dragonfly Geophysics and Meteorology) package was specified in order to measure the boundary layer profile during flights up to 3-4 km.  The  absolute accuracy specification is therefore only about 2 mbar  (200 Pa).  However, the sensitivity to changes will be considerably better, and the telemetered resolution may be as good as 1 Pa. The precision that can be practically achieved (and thus the ability to detect small changes) will, as on other missions, be contingent on sensor noise performance and will likely be period-dependent and lander-activity-dependent: tests in Phase A with representative sensors suggest that the ability to detect variations with an amplitude of 5 Pa on a range of periods should be possible with suitable filtering. 
As discussed in \cite{chapman1970}, periodic changes can be detected in long time series data by virtue of averaging over many cycles. The Dragonfly nominal mission is 3 years, or about 70 Tsols (Titan solar days). Crudely, diurnal signals may be measurable with a precision of $\sim$70$^{0.5}$ $\sim$8$\times$ better than that of individual measurements, offering good prospects that an atmospheric tidal signal can be measured with a precision that can discriminate interior structure models. 
For comparison, the pressure sensor on the Huygens probe had an accuracy lower than 100 Pa and a resolution lower 1 Pa, quite similar to Dragonfly's sensors. The Huygens probe measured surface pressure on Titan for around 30 minutes \citep{harri2006}. The standard deviation of those pressure readings was $\sim$1 Pa, indicating that any variation in background pressure must have been slower than 2 Pa/hour. Unfortunately the measurement was not long enough to constrain the tides. The maximal pressure variation (for the ideal case with no deformation of the interior) due to tides at the Huygens site is $\frac{dP}{dt}=-\Omega \rho_0 V_0 \sim$ 1 Pa/hour. However, the precision of Huygen's measurements holds promise for Dragonfly's ability to detect tides.

It is probable that the largest confounding factor in retrieving a tidal pressure signal is that lander activities will be correlated with the time of (solar) day. Notably, rotorcraft flight and communication with Earth, both of which will entail appreciable energy dissipation in the lander, and thus some change in internal temperatures including the pressure sensor, will occur during daytime. Thus any analysis of the sort described here must be vigilant towards imperfections in the temperature compensation of the pressure measurement (a known challenge on other missions - e.g.  \cite{taylor2010})  which may lead to a spurious signal with a period of one Tsol (i.e. the same period as the atmospheric tide), and it is likely that the magnitude of such effects will not be reliably known until Titan arrival.   Over three years of landed operations, however (a tenth of a Saturnian year), the phasing of the gravitational tidal peak, fixed with respect to Titan's orbit around Saturn and thus in inertial space, will change with respect to local solar time   (due to the difference in length of the solar and sidereal day), which will mitigate this problem.  It should also be recognized that while it is intended that DraGMet measure pressure data many times per Titan day (perhaps Earth-hourly, with some bursts of higher-rate data), the record will have step-like jumps in pressure owing to the different elevations of various landing sites.  Thus some manual editing and adjustment preprocessing of the time series may be necessary to most precisely extract a tidal signal. 

\subsection{Constraints on Titan's interior}

As discussed in Section 2, the degree-2 Love numbers ($k_2$ and $h_2$) and the surface pressure variations are related together by:
\begin{equation}
\begin{split} 
& P_{\rm surf}=-\rho_0 V_1 (1+ \Re(k_2-h_2))\cos(\Omega t-\psi) \\
& + \rho_0 V_1 \Im(k_2-h_2)\sin(\Omega t-\psi) 
\end{split} 
\label{eq9}
\end{equation}
Assuming that Dragonfly's pressure measurements have a gaussian noise with an uncertainty equal to $\sigma_P$,
the uncertainty on the tidal pressure amplitude would be
equal to $\sqrt{\frac{2}{N_{obs}}}\sigma_P$ and the uncertainty on $1+\Re(k_2-h_2)$ and $\Im(k_2-h_2)$ would be equal to $\sqrt{\frac{2}{N_{obs}}}\frac{\sigma_P}{\rho_0 V_1}$, where $N_{obs}$ is the number of measurements \citep{alegria2006}.
Table 1 shows the uncertainty for these parameters after 1 or 70 Titan days, assuming a precision of either 50 Pa (optimistic case) or 200 Pa (pessimistic case) for each individual measurement, done every hour (so 383 measurements per Tsol). The pessimistic case corresponds to the absolute accuracy specification. We have chosen a precision four times better for the optimistic case, able to detect a variation of 4 Pa amplitude after one Tsol. This is fairly consistent with initial tests of Dragonfly's pressure sensors.
Based on these two scenarios, we expect that Dragonfly could detect tidal pressure variations and could
measure the real and imaginary parts of $(k_2-h_2)$ with a precision of $\pm$ 0.01-0.03 over the whole mission.
This precision is comparable to the precision on $k_2$ from Cassini flybys and will be complementary to the $k_2$ estimate.
According to Fig. \ref{figure_2}, such values would allow us to derive the ice shell thickness with a precision of around $\pm$ 15 km and to estimate the heat flux with a precision of around $\pm$ 5 mW/m$^2$, using interior models. Unfortunately, the determination of the ocean density would be limited by the precision on $k_2$. Some values of density could however be excluded. For instance 1+ $\Re(k_2-h_2)$ $\ge$ 0.08 would imply a density higher than 1250 kg.m$^{-3}$.

\begin{table}[h]
\begin{center}
\begin{tabular}{|c|c|c|c|c|}
  \hline
  Case & \multicolumn{2}{c|}{Optimistic}  & \multicolumn{2}{c|}{Pessimistic} \\ \hline 
  Number of Tsols & 1  & 70  & 1 & 70 \\
  \hline
  Amplitude (Pa) & $\pm$4   & $\pm$0.4  & $\pm$14 & $\pm$1.7  \\
 $\Re(h_2-k_2$)        & $\pm$0.06 & $\pm$0.007 & $\pm$0.2 & $\pm$0.03 \\
  $\Im(h_2-k_2$)    & $\pm$0.06 & $\pm$0.007 & $\pm$0.2 & $\pm$0.03 \\  
  \hline
\end{tabular}
\end{center}
\caption{Expected precision of Dragonfly measurements for the amplitude of tidal pressure variations, for the real and the imaginary parts of $k_2-h_2$. Values are computed for an optimistic case (precision of 50 Pa for each individual measurement) and a pessimistic case (precision of 200 Pa for each individual measurement). They are given for 1 Titan day or 70 Titan days (i.e. the whole mission), assuming measurements every hour (383 measurements per Titan day).}
\end{table}

\section{Summary and conclusions}

In this article, we reanalysed Titan's gravitational atmospheric tides, including the effects of the deformation of the interior. We showed that the tidal response of the interior, characterized by low values of $1+\Re(k_2-h_2)$ and $\Im(k_2-h_2)$, should strongly decrease the tidal potential affecting the atmosphere. Using analytical calculations and 3D GCM simulations, we showed that the atmosphere should quickly respond to the tidal potential, with almost no phase shift and extremely weak tidal winds. We predict that the amplitude of the pressure variation should be $\sim$5 Pa. This low value mitigates the possible effect of gravitational tides for transporting energy in the upper atmosphere, as suggested by \cite{strobel2006}. In addition, the impact of gravitational tides on cloud formation should be totally negligible. Yet the tidal pressure variations could be higher than the variations caused by thermal tides or planetary waves in the equatorial region. The methane cycle (i.e. tropospheric cloud formation and precipitation), which was not included in our 3D simulations, could be an additional source of variability.

Finally, we predict that Dragonfly can certainly detect the pressure signal that would be present if tidal deformation of the interior did not occur. It may detect even the small residual tidal pressure variations taking interior/crustal deformation into account. The real and imaginary part of $(k_2-h_2)$ could then be inferred with a precision of $\pm$ 0.01-0.03 over the whole mission, constraining the thickness of the ice shell and (via models) the internal heat flux with a precision of $\sim$ $\pm$ 15 km and $\sim$ $\pm$ 5 mW/m$^2$ respectively. 
These measurements are insensitive to the ocean density, but can restrict the range of possible ocean density in combination with current $k_2$ estimate. A future orbiter around Titan could strongly improve the estimate of $k_2$. Combined with Dragonfly measurements, it could then constrain the density of the internal ocean, with implications on its composition, evolution and habitability.

\begin{acknowledgements}
This work was granted access to the HPC resources of MesoPSL financed by the Region Ile de France and the project Equip@Meso (reference ANR-10-EQPX-29-01) of the programme Investissements d’Avenir supervised by the Agence Nationale pour la Recherche. This work was supported by the Programme National de Planétologie (PNP) of CNRS/INSU, co-funded by CNES. GT acknowledges the support from the ANR COLOSSe project.
RL acknowledges the support of the Dragonfly project, via NASA contract NNN06AA01C to the Johns Hopkins Applied Physics Laboratory. 
We are grateful to Henrik Kahanp\"a\"a for a useful discussion about Huygens' pressure measurements, and we thank the referee, Tetsuya Tokano, for his thorough review.

\end{acknowledgements}

\newpage
\bibliographystyle{apj}
\bibliography{biblio_total}

\begin{thebibliography}{}
\expandafter\ifx\csname natexlab\endcsname\relax\def\natexlab#1{#1}\fi

\bibitem[{Alegria \& Serra(2006)}]{alegria2006}
Alegria, F.~C., \& Serra, A.~C. 2006, in 2006 IEEE Instrumentation and
  Measurement Technology Conference Proceedings, 1643--1647

\bibitem[{Barnes(2021)}]{barnes2021}
Barnes, J. e.~a. 2021, The Planetary Science Journal, 2, 18

\bibitem[{Castillo-Rogez {et~al.}(2011)Castillo-Rogez, Efroimsky, \&
  Lainey}]{castillo2011}
Castillo-Rogez, J.~C., Efroimsky, M., \& Lainey, V. 2011, Journal of
  Geophysical Research: Planets, 116

\bibitem[{Castillo-Rogez \& Lunine(2010)}]{castillo2010}
Castillo-Rogez, J.~C., \& Lunine, J.~I. 2010, Geophysical Research Letters, 37

\bibitem[{Chapman \& Lindzen(1970)}]{chapman1970}
Chapman, S., \& Lindzen, R. 1970, Atmospheric Tides (D. Reidel Publ. Co.,
  Dordrecht, Holland), 200

\bibitem[{Charnay \& Lebonnois(2012)}]{charnay2012}
Charnay, B., \& Lebonnois, S. 2012, Nature Geoscience, 5, 106

\bibitem[{Dumoulin {et~al.}(1999)Dumoulin, Doin, \& Fleitout}]{dumoulin1999}
Dumoulin, C., Doin, M.-P., \& Fleitout, L. 1999, Journal of Geophysical
  Research: Solid Earth, 104, 12759

\bibitem[{{Durante} {et~al.}(2019){Durante}, {Hemingway}, {Racioppa}, {Iess},
  \& {Stevenson}}]{durante2019}
{Durante}, D., {Hemingway}, D.~J., {Racioppa}, P., {Iess}, L., \& {Stevenson},
  D.~J. 2019, Icarus, 326, 123

\bibitem[{{Friedson} {et~al.}(2009){Friedson}, {West}, {Wilson}, {Oyafuso}, \&
  {Orton}}]{friedson2009}
{Friedson}, A.~J., {West}, R.~A., {Wilson}, E.~H., {Oyafuso}, F., \& {Orton},
  G.~S. 2009, \planss, 57, 1931

\bibitem[{{Harri} {et~al.}(2006){Harri}, {M{\"a}kinen}, {Lehto},
  {Kahanp{\"a}{\"a}}, \& {Siili}}]{harri2006}
{Harri}, A.-M., {M{\"a}kinen}, T., {Lehto}, A., {Kahanp{\"a}{\"a}}, H., \&
  {Siili}, T. 2006, \planss, 54, 1117

\bibitem[{Iess {et~al.}(2010)Iess, Rappaport, Jacobson, Racioppa, Stevenson,
  Tortora, Armstrong, \& Asmar}]{iess2010}
Iess, L., Rappaport, N.~J., Jacobson, R.~A., {et~al.} 2010, science, 327, 1367

\bibitem[{{Iess} {et~al.}(2012){Iess}, {Jacobson}, {Ducci}, {Stevenson},
  {Lunine}, {Armstrong}, {Asmar}, {Racioppa}, {Rappaport}, \&
  {Tortora}}]{iess2012}
{Iess}, L., {Jacobson}, R.~A., {Ducci}, M., {et~al.} 2012, Science, 337, 457

\bibitem[{Journaux {et~al.}(2020)Journaux, Kalousov{\'a}, Sotin, Tobie, Vance,
  Saur, Bollengier, Noack, R{\"u}ckriemen-Bez, Van~Hoolst,
  {et~al.}}]{journaux2020}
Journaux, B., Kalousov{\'a}, K., Sotin, C., {et~al.} 2020, Space Science
  Reviews, 216, 1

\bibitem[{{Kohyama} \& {Wallace}(2016)}]{kohyama2016}
{Kohyama}, T., \& {Wallace}, J.~M. 2016, \grl, 43, 918

\bibitem[{{Lebonnois} {et~al.}(2012){Lebonnois}, {Burgalat}, {Rannou}, \&
  {Charnay}}]{lebonnois2012}
{Lebonnois}, S., {Burgalat}, J., {Rannou}, P., \& {Charnay}, B. 2012, Icarus,
  218, 707

\bibitem[{Lefevre {et~al.}(2014)Lefevre, Tobie, Choblet, \&
  {\v{C}}adek}]{lefevre2014}
Lefevre, A., Tobie, G., Choblet, G., \& {\v{C}}adek, O. 2014, Icarus, 237, 16

\bibitem[{Lorenz {et~al.}(2019)Lorenz, Panning, Stahler, Shiraishi, Yamada, \&
  Turtle}]{lorenz2019}
Lorenz, R., Panning, M., Stahler, S., {et~al.} 2019, in Lunar and Planetary
  Science Conference, Vol.~50, 2173

\bibitem[{Lorenz(1992)}]{lorenz1992}
Lorenz, R.~D. 1992, in Proceedings of the Symposium on Titan, Toulouse,
  September 1991, 119--123

\bibitem[{Lorenz \& Le~Gall(2020)}]{lorenz2020}
Lorenz, R.~D., \& Le~Gall, A. 2020, Icarus, 351, 113942

\bibitem[{{Lorenz} {et~al.}(2014){Lorenz}, {Kirk}, {Hayes}, {Anderson},
  {Lunine}, {Tokano}, {Turtle}, {Malaska}, {Soderblom}, {Lucas}, {Karatekin},
  \& {Wall}}]{lorenz2014}
{Lorenz}, R.~D., {Kirk}, R.~L., {Hayes}, A.~G., {et~al.} 2014, Icarus, 237, 9

\bibitem[{Lorenz {et~al.}(2018)Lorenz, Turtle, Barnes, Trainer, Adams, Hibbard,
  Sheldon, Zacny, Peplowski, Lawrence, {et~al.}}]{lorenz2018}
Lorenz, R.~D., Turtle, E.~P., Barnes, J.~W., {et~al.} 2018, Johns Hopkins APL
  Technical Digest

\bibitem[{Lorenz {et~al.}(2021)Lorenz, MacKenzie, Neish, Le~Gall, Turtle,
  Barnes, Trainer, Werynski, Hedgepeth, \& Karkoschka}]{lorenz2021}
Lorenz, R.~D., MacKenzie, S.~M., Neish, C.~D., {et~al.} 2021, The Planetary
  Science Journal, 2, 24

\bibitem[{{Mitchell} {et~al.}(2011){Mitchell}, {{\'A}d{\'a}mkovics},
  {Caballero}, \& {Turtle}}]{mitchell2011}
{Mitchell}, J.~L., {{\'A}d{\'a}mkovics}, M., {Caballero}, R., \& {Turtle},
  E.~P. 2011, Nature Geoscience, 4, 589

\bibitem[{Mitri {et~al.}(2014)Mitri, Meriggiola, Hayes, Lefevre, Tobie, Genova,
  Lunine, \& Zebker}]{mitri2014}
Mitri, G., Meriggiola, R., Hayes, A., {et~al.} 2014, Icarus, 236, 169

\bibitem[{N{\'e}ri {et~al.}(2020)N{\'e}ri, Guyot, Reynard, \& Sotin}]{neri2020}
N{\'e}ri, A., Guyot, F., Reynard, B., \& Sotin, C. 2020, Earth and Planetary
  Science Letters, 530, 115920

\bibitem[{{Rodriguez} {et~al.}(2009){Rodriguez}, {Le Mou{\'e}lic}, {Rannou},
  {Tobie}, {Baines}, {Barnes}, {Griffith}, {Hirtzig}, {Pitman}, {Sotin},
  {Brown}, {Buratti}, {Clark}, \& {Nicholson}}]{rodriguez2009}
{Rodriguez}, S., {Le Mou{\'e}lic}, S., {Rannou}, P., {et~al.} 2009, \nat, 459,
  678

\bibitem[{{Sagan} \& {Dermott}(1982)}]{sagan1982}
{Sagan}, C., \& {Dermott}, S.~F. 1982, \nat, 300, 731

\bibitem[{Saito(1974)}]{saito1974}
Saito, M. 1974, Journal of Physics of the Earth, 22, 123

\bibitem[{{Sohl} {et~al.}(2003){Sohl}, {Hussmann}, {Schwentker}, {Spohn}, \&
  {Lorenz}}]{sohl2003}
{Sohl}, F., {Hussmann}, H., {Schwentker}, B., {Spohn}, T., \& {Lorenz}, R.~D.
  2003, Journal of Geophysical Research (Planets), 108, 5130

\bibitem[{{Sohl} {et~al.}(1995){Sohl}, {Sears}, \& {Lorenz}}]{sohl1995}
{Sohl}, F., {Sears}, W.~D., \& {Lorenz}, R.~D. 1995, \icarus, 115, 278

\bibitem[{Sotin {et~al.}(2021)Sotin, Kalousová, \& Tobie}]{sotin2021}
Sotin, C., Kalousová, K., \& Tobie, G. 2021, Annual Review of Earth and
  Planetary Sciences, 49, 579

\bibitem[{{Strobel}(2006)}]{strobel2006}
{Strobel}, D.~F. 2006, Icarus, 182, 251

\bibitem[{{Takeuchi} \& {Saito}(1972)}]{takeuchi1972}
{Takeuchi}, H., \& {Saito}, M. 1972, in Methods in Computational Physics:
  Advances in Research and Applications, Vol.~11, Seismology: Surface Waves and
  Earth Oscillations, ed. B.~A. BOLT (Elsevier), 217--295

\bibitem[{{Taylor} {et~al.}(2010){Taylor}, {Kahanp{\"a}{\"a}}, {Weng},
  {Akingunola}, {Cook}, {Daly}, {Dickinson}, {Harri}, {Hill}, {Hipkin},
  {Polkko}, \& {Whiteway}}]{taylor2010}
{Taylor}, P.~A., {Kahanp{\"a}{\"a}}, H., {Weng}, W., {et~al.} 2010, Journal of
  Geophysical Research (Planets), 115, E00E15

\bibitem[{Tobie {et~al.}(2005)Tobie, Mocquet, \& Sotin}]{tobie2005}
Tobie, G., Mocquet, A., \& Sotin, C. 2005, Icarus, 177, 534

\bibitem[{{Tokano}(2010{\natexlab{a}})}]{tokano2010b}
{Tokano}, T. 2010{\natexlab{a}}, Ocean Dynamics, 60, 803

\bibitem[{{Tokano}(2010{\natexlab{b}})}]{tokano2010}
---. 2010{\natexlab{b}}, Planetary and Space Science, 58, 814

\bibitem[{{Tokano} {et~al.}(2014){Tokano}, {Lorenz}, \& {Van
  Hoolst}}]{tokano2014}
{Tokano}, T., {Lorenz}, R.~D., \& {Van Hoolst}, T. 2014, Icarus, 242, 188

\bibitem[{{Tokano} \& {Neubauer}(2002)}]{tokano2002}
{Tokano}, T., \& {Neubauer}, F.~M. 2002, Icarus, 158, 499

\bibitem[{{Turtle} {et~al.}(2011){Turtle}, {Del Genio}, {Barbara}, {Perry},
  {Schaller}, {McEwen}, {West}, \& {Ray}}]{turtle2011}
{Turtle}, E.~P., {Del Genio}, A.~D., {Barbara}, J.~M., {et~al.} 2011, \grl, 38,
  L03203

\bibitem[{Turtle {et~al.}(2018)Turtle, Barnes, Trainer, Lorenz, Hibbard, Adams,
  Bedini, Brinckerhoff, Cable, Ernst, {et~al.}}]{turtle2018}
Turtle, E.~P., Barnes, J., Trainer, M., {et~al.} 2018, in Lunar and {{Planetary
  Science Conference}}, Vol.~49

\bibitem[{{Tyler}(2008)}]{tyler2008}
{Tyler}, R.~H. 2008, \nat, 456, 770

\bibitem[{{Vincent} {et~al.}(2018){Vincent}, {Karatekin}, {Lambrechts},
  {Lorenz}, {Dehant}, \& {Deleersnijder}}]{vincent2018}
{Vincent}, D., {Karatekin}, {\"O}., {Lambrechts}, J., {et~al.} 2018, \icarus,
  310, 105

\bibitem[{{Vincent} {et~al.}(2016){Vincent}, {Karatekin}, {Vallaeys}, {Hayes},
  {Mastrogiuseppe}, {Notarnicola}, {Dehant}, \& {Deleersnijder}}]{vincent2016}
{Vincent}, D., {Karatekin}, {\"O}., {Vallaeys}, V., {et~al.} 2016, Ocean
  Dynamics, 66, 461

\bibitem[{{Walterscheid} \& {Schubert}(2006)}]{walterscheid2006}
{Walterscheid}, R.~L., \& {Schubert}, G. 2006, \icarus, 183, 471

\bibitem[{{Wang} {et~al.}(2016){Wang}, {Boyd}, \& {Akmaev}}]{wang2016}
{Wang}, H., {Boyd}, J.~P., \& {Akmaev}, R.~A. 2016, Geoscientific Model
  Development, 9, 1477

\end{thebibliography}

\end{document}